\documentclass[final,3p,times]{elsarticle}
\usepackage[latin9]{inputenc}
\usepackage{bm}
\usepackage{amsmath}
\usepackage{amssymb}
\usepackage{graphicx}
\usepackage{esint}
\usepackage{multirow}
\usepackage[subfigure]{graphfig}
\usepackage{rotating}
\usepackage{verbatim}
\usepackage{setspace}
\usepackage[dvipsnames]{xcolor}
\makeatletter
\newcommand*{\rom}[1]{\expandafter\@slowromancap\romannumeral #1@}

\usepackage{tabularx}\usepackage{subfigure}\usepackage{subfigure}\usepackage{color}

\usepackage{setspace}

\usepackage{bm}

\newfont{\numerikEleven}{ecrm1000}
\newfont{\numerikTen}{cmss10}
\newfont{\numerikNine}{cmss9}
\newfont{\numerikEight}{cmss8}
\newfont{\numerikSeven}{cmss7}
\newfont{\numerikSix}{cmss6}

\journal{}

\makeatother

\color{black}

\begin{document}

\title{A fifth-order shock capturing scheme with cascade boundary variation diminishing algorithm}

\author[ad1]{Xi Deng \corref{cor}}

\author[ad1]{Yuya Shimizu}

\author[ad1]{Feng Xiao \corref{cor} }


\address[ad1]{Department of Mechanical Engineering, Tokyo Institute of Technology, 2-12-1 Ookayama, Meguro-ku, Tokyo, 152-8550, Japan.}

\cortext[cor]{Corresponding author: 
Mr. X. Deng (Email: deng.xi98@gmail.com), Dr. F. Xiao (xiao.f.aa@m.titech.ac.jp)
}

\begin{abstract}
A novel 5th-order shock capturing scheme is presented in this paper. The scheme, so-called  P4-THINC-BVD  (4th degree polynomial and THINC reconstruction based on BVD algorithm), is formulated as a two-stage cascade BVD (Boundary Variation Diminishing) algorithm following the BVD principle that minimizes the jumps of reconstructed values at cell boundaries.  In the P4-THINC-BVD scheme, polynomial of degree four and THINC (Tangent of Hyperbola for INterface Capturing) functions with adaptive steepness are used as the candidate reconstruction functions. The final reconstruction function is selected from the candidate functions by a two-stage cascade BVD algorithm so as to effectively control numerical oscillation and dissipation. Spectral analysis and numerical verifications show that the P4-THINC-BVD scheme possesses the following desirable properties: 1) it effectively suppresses spurious numerical oscillation in the presence of strong shock or discontinuity; 2) it substantially reduces numerical dissipation errors; 3) it automatically retrieves the underlying linear 5th-order upwind scheme for smooth solution over all wave numbers;  4) it is able to resolve both smooth and discontinuous flow structures of all scales with substantially improved solution quality in comparison to other existing methods; and 5) it faithfully maintains the  free-mode solutions in long term computation. P4-THINC-BVD, as well as the underlying idea presented in this paper, provides an innovative and  practical approach to design high-fidelity numerical schemes  for compressible flows involving strong discontinuities and flow structures of wide range scales. 

\end{abstract}

\maketitle

\section{Introduction}
Developing shock capturing schemes for compressible flows involving both discontinuous and smooth solutions still remains a challenging problem to computational fluid dynamics community. The coexistence of discontinuities and smooth structures of different scales poses difficulties to existing numerical schemes. On the one hand, high order schemes with less numerical dissipation  are demanded to resolve structures in vortical and turbulent flows. On the other hand, high order schemes may cause spurious numerical oscillations when solving discontinuities such as shock waves, contacts and material interfaces. The common practice to suppress numerical oscillations associated with high order reconstruction is to introduce certain amount of numerical dissipations by projecting high order interpolation polynomials to lower order or smoother ones, which is also known as limiting projection. An ideal limiting projection is expected to introduce as small as possible numerical dissipations to smooth solutions while sufficiently suppressing spurious numerical oscillations around discontinuous or large gradient regions. Unfortunately, the current limiting projection methods  which are based on polynomial form and the compromise between numerical oscillation and numerical dissipation cannot satisfactorily resolve both smooth and non-smooth solutions. Thus, new methods of better fidelity are highly demanded in the simulations of high-Mach compressible flow and interfacial multiphase flow where both discontinuous and vortical structures are dominant.      

Over the decades, a great deal of efforts have been made to construct shock capturing schemes. TVD (Total Variation Diminishing) schemes \cite{harten-tvd,sweby84}, or equivalently the MUSCL (Monotone Upstream-centered Schemes for Conservation Law) schemes \cite{Van_Leer}, can effectively eliminate numerical oscillations by introducing flux or slope limiters which are solution dependent and thus lead to a class of nonlinear schemes widely used in simulating flows with discontinuous solutions. However, TVD schemes, which can be of 2nd-order accuracy in 1D at most,  suffer from excessive numerical dissipation and smear out flow structures, like vortices and contact discontinuities. 

A remedy to numerical dissipation is using higher order polynomials for smooth solutions, and ENO (Essentially Non-oscillatory) scheme  \cite{Harten1,Harten2,shu_eno1,shu_eno2} has been proposed to suppress numerical oscillation where high-order polynomials are used. The basic idea of ENO is to construct several polynomial interpolations over different candidate stencils and to choose the smoothest one as the final reconstruction function, where a numerical formula, so-called smoothness indicator,  is devised to quantify the smoothness of each interpolation polynomial. Following the ENO schemes, WENO (Weighted Essentially Non-oscillatory) schemes have been devised in \cite{liu94,jiang96}. Instead of choosing the smoothest stencil, WENO reconstructions are built on a weighted average of polynomial approximations over all candidate stencils. The weight of each stencil is assigned according to smoothness indicators. In the vicinity of discontinuities, ENO property is recovered through assigning small weights to less smooth candidates. In smooth regions, weights are designed to restore the underlying linear schemes. With more candidate polynomials involved, WENO schemes are more accurate in smooth region than ENO schemes. As shown in \cite{jiang96}, the WENO scheme achieves expected high order accuracy for smooth solution while keeps the  essentially non-oscillatory property near discontinuities. The underlying idea of WENO indicates a practical strategy to design  schemes that can be of arbitrarily high order yet effectively suppress numerical oscillation to an extent less restrictive than the TVD criterion. Hence, WENO scheme substantially reduces numerical dissipation compared to the TVD schemes. However, WENO scheme is still found to be too dissipative in some applications, as addressed in \cite{WENOM,TENO14,TENO15} for example,  WENO scheme generates excessive numerical dissipation that tends to smear out contact discontinuities or jumps across material interfaces, as well as the small-scale flow structures in  simulations including acoustic effect or turbulence.  In successive works following  \cite{liu94,jiang96}, efforts have been made  to reduce numerical dissipation. Variants of the smoothness indicator have been proposed in \cite{wenoz,wenoh,wenop,wenozn} where contributions of the less smooth candidate stencils are optimized to reduce numerical dissipation. Being a particular tuning for simulation of turbulent flow, numerical dissipation can be further reduced by  employing central discretization \cite{wenohu}. It is observed that the spectral property of the WENO scheme is inferior to that of the corresponding linear scheme. In \cite{TENO}, a family of high order targeted ENO (TENO) schemes are proposed to restore the low-dissipation feature of the linear schemes. In \cite{embedded}, a new design strategy named embedded WENO is devised to utilize all adjacent smooth substencils to construct the interpolation so as to improve the spectral property. All these works attempt to decrease the effect of nonlinear WENO adaptation and to retrieve the low-dissipation linear scheme as much as  possible.  The present work demonstrates another path to make the largest  possible use of high-order linear schemes for spatial reconstruction.       

A novel guideline for constructing high-fidelity shock capturing schemes with small numerical dissipation, so-called  BVD (Boundary Variation Diminishing), was proposed in  \cite{Sun} for general finite volume method. The BVD principle requires that the jumps of the reconstructed values at cell boundaries should be minimized so as to reduce the effective numerical dissipation in Riemann solvers. In \cite{Sun}, a hybrid reconstruction of WENO and THINC (Tangent of Hyperbola for INterface Capturing) is designed with a BVD algorithm which automatically switches to the WENO reconstruction for smooth region and to the THINC reconstruction for discontinuous solution. The resulting scheme, WENO-THINC-BVD, maintains the accuracy (convergence rate) of the WENO scheme for smooth solution, while substantially improves the solution quality for discontinuity. Since then, several BVD variant algorithms and other BVD-admissible reconstruction functions have been devised and tested to facilitate the implementation of the BVD principle\cite{xie2017,deng2018a,deng2018b}.  It is found that the BVD principle with THINC as one of the candidate reconstruction schemes provides a framework to design schemes that are able to resolve both smooth and discontinuous solutions with superior  accuracy to the existing methods. It also indicates the possibility and necessity to explore new methods with better numerical properties in spirit of the BVD principle. 

In the present work, we propose a new BVD reconstruction strategy using the  linear 5th order upwind scheme of 4th degree polynomial as one of the candidate reconstruction functions to enjoy the advantages of the linear schemes for smooth solutions.  The new method, so-called P4-THINC-BVD (4th-degree polynomial and THINC reconstruction based on BVD algorithm) scheme makes use of a two-stage cascade BVD algorithm to get high order and oscillation-less numrical results.  In the first stage, the 4th-degree polynomial and a THINC function with a moderate  steepness are used as the candidate reconstruction functions in the BVD framework, which results in oscillation-less numerical results. {The second stage is devised to reduce numerical dissipation associated with less smooth solutions, where sharp discontinuities are captured with a steeper THINC function.} The spectral analysis shows that the new scheme can restore the underlying low-dissipation linear scheme even for high wavenumber band. The superior capability of P4-THINC-BVD in capturing sharp discontinuity and resolving small-scale flow structures with low-dissipation has been verified through widely used benchmark tests. 


The remainder of this paper is organized as follows. In Section 2, after a brief review of the finite volume method, the details of the new scheme for spatial reconstruction are presented. The spectral property of the new scheme will also be presented. In Section 3, the performance of the new scheme will be examined through benchmark tests in 1D and 2D with comparison with the 5th-order WENO schemes. Some concluding remarks are given in Section 4. 

\section{Numerical methods \label{sec:model}}
We use  the 1D scalar conservation law in the following form to introduce the new scheme
\begin{equation}
\label{eq:scalar}
\frac{\partial q}{\partial t} + \frac{\partial f(q)}{\partial x} = 0,
\end{equation}
where $q(x,t)$ is the solution function and $f(q)$ is the flux function. We assume that the flux function has hyperbolicity, i.e. $\partial f(q)/\partial q=a$, the  characteristic speed,  is a real number. 
\subsection{Finite volume method}
 We divide the computational domain into $N$ non-overlapping cell elements, ${\mathcal I}_{i}: x \in [x_{i-1/2},x_{i+1/2} ]$, $i=1,2,\ldots,N$, with a uniform grid spacing $h=\Delta x=x_{i+1/2}-x_{i-1/2}$. For a standard finite volume method, the volume-integrated average value $\bar{q}_{i}(t)$ in cell ${\mathcal I}_{i}$ is defined as
\begin{equation}
\bar{q}_{i}(t) 
\approx \frac{1}{\Delta x} \int_{x_{i-1/2}}^{x_{i+1/2}}
q(x,t) \; dx.
\end{equation}
The semi-discrete version of Eq.~(\ref{eq:scalar}) in the finite volume form for cell ${\mathcal I}_{i}$ can be expressed as an ordinary differential equation (ODE)
\begin{equation}
\frac{\partial \bar{q}_i(t)}{\partial t}  =-\frac{1}{\Delta x}(\tilde{f}_{i+1/2}-\tilde{f}_{i-1/2}),
\end{equation}
where the numerical fluxes $\tilde{f}$ at cell boundaries can be computed by a Riemann solver
\begin{equation}
\tilde{f}_{i+1/2}=f_{i+1/2}^{\text{Riemann}}(q_{i+1/2}^{L},q_{i+1/2}^{R}), 
\end{equation}
once the reconstructed left-side value $q_{i+1/2}^{L}$ and right-side value $q_{i+1/2}^{R}$ at cell boundaries are provided. Essentially, the Riemann flux can be written into a canonical form as
\begin{equation}
\label{eq:Riemann}
f_{i+1/2}^{\text{Riemann}}(q_{i+1/2}^{L},q_{i+1/2}^{R}) =\frac{1}{2}\left(f(q_{i+1/2}^{L})+f(q_{i+1/2}^{R})\right)-\frac{|a_{i+1/2}|}{2}\left(q_{i+1/2}^{R}-q_{i+1/2}^{L}\right),
\end{equation}
where $a_{i+1/2}$ stands for the characteristic speed of the hyperbolic conservation law. The remaining main task is how to calculate $q_{i+1/2}^{L}$ and $q_{i+1/2}^{R}$ through the reconstruction process.

\subsection{Reconstruction process}
In this subsection, we describe the details of how to calculate $q_{i+1/2}^{L}$ and $q_{i+1/2}^{R}$ using the BVD principle. This new scheme, P4-THINC-BVD scheme, is designed by using a cascade BVD algorithm where a linear upwind scheme of 4th degree polynomial and a THINC function are employed as the candidate interpolants to get high fidelity solutions for both smooth and discontinuous solutions. Next, we introduce the candidate interpolants before the description of the BVD algorithm. 

\subsubsection{Candidate interpolant 1: linear upwind scheme of 4th degree polynomial}
A finite volume scheme of $(D+1)$th order can be constructed from a spatial approximation for the solution in the target cell ${\mathcal I}_{i}$ with a polynomial $\tilde{q}_{i}^{PD}(x)$ of degree $D$. The $D+1$ unknown coefficients of the polynomial are determined by requiring that $\tilde{q}_{i}^{PD}(x)$ has the same cell averages on each cell over an appropriately selected stencil $S=\{i-D^{-},\dots, i+D^{+}\}$ with $D^{-}+D^{+}=D$, which is expressed as
\begin{equation}\label{Eq:linearR}
\dfrac{1}{\Delta x}\int_{x_{j-1/2}}^{x_{j+1/2}}\tilde{q}_{i}^{PD}(x)dx= \bar{q}_{j}, ~~j=i-D^{-},i-D^{-}+1,\dots,i+D^{+}.
\end{equation}
To construct $2r-1$ order upwind-biased finite volume schemes as detailed in \cite{liu94,jiang96,very3}, the stencil is defined with $D^{-}=D^{+}=r-1$. The unknown coefficients of polynomial of $2r-2$ degree can be then calculated from \eqref{Eq:linearR}. With the polynomial $\tilde{q}_{i}^{PD}(x)$, high order approximation for reconstructed values at the cell boundaries can be obtained by
\begin{equation}\label{Eq:upwind}
q^{L,PD}_{i+\frac{1}{2}}=\tilde{q}_{i}^{PD}(x_{i+\frac{1}{2}}) \ \ {\rm and} \ \  q^{R,PD}_{i-\frac{1}{2}}=\tilde{q}_{i}^{PD}(x_{i-\frac{1}{2}}).
\end{equation}
The analysis of \cite{liu94,jiang96,very3} shows that in smooth region, the approximation with polynomial \eqref{Eq:upwind} can achieve $2r-1$ order accuracy. In this work, we use the fifth order ($r=3$) upwind-biased scheme as the underlying scheme for smooth solution. We denote the upwind scheme with polynomial of degree four reconstruction function in cell ${\mathcal I}_{i}$ by $\tilde{q}_{i}^{P4}(x)$ and the reconstructed value at cell boundaries as $q_{i+1/2}^{L,P4}$ and $q_{i-1/2}^{R,P4}$. The explicit formulas of $q_{i+1/2}^{L,P4}$ and $q_{i-1/2}^{R,P4}$ are given by 
\begin{equation}\label{Eq:5thuw}
\begin{aligned}
q_{i+1/2}^{L,P4}=\dfrac{1}{30}\bar{q}_{i-2}-\dfrac{13}{60}\bar{q}_{i-1}+\dfrac{47}{60}\bar{q}_{i}+\dfrac{9}{20}\bar{q}_{i+1}-\dfrac{1}{20}\bar{q}_{i+2}, \\
q_{i-1/2}^{R,P4}=\dfrac{1}{30}\bar{q}_{i+2}-\dfrac{13}{60}\bar{q}_{i+1}+\dfrac{47}{60}\bar{q}_{i}+\dfrac{9}{20}\bar{q}_{i-1}-\dfrac{1}{20}\bar{q}_{i-2}.
\end{aligned}
\end{equation}

It is noted that beyond the 5th order upwind scheme used in this work, polynomials of higher degree can also be used alternatively as the candidate interpolants to get higher order convergence rates for smooth solution following the same spirit.

\subsubsection{Candidate interpolant 2: non-polynomial THINC function}
Another candidate interpolation function in our scheme makes use of the THINC interpolation which is a differentiable and monotone Sigmoid function \cite{xiao_thinc,xiao_thinc2}.  
The piecewise THINC reconstruction function is written as
\begin{equation} \label{eq:THINC}
\tilde{q}_{i}^{T}(x)=\bar{q}_{min}+\dfrac{\bar{q}_{max}}{2} \left[1+\theta~\tanh \left(\beta \left(\dfrac{x-x_{i-1/2}}{x_{i+1/2}-x_{i-1/2}}-\tilde{x}_{i}\right)\right)\right],
\end{equation} 
where $\bar{q}_{min}=\min(\bar{q}_{i-1},\bar{q}_{i+1})$, $\bar{q}_{max}=\max(\bar{q}_{i-1},\bar{q}_{i+1})-\bar{q}_{min}$ and 
$\theta=sgn(\bar{q}_{i+1}-\bar{q}_{i-1})$. The jump thickness is controlled by the parameter $\beta$, i.e. a small value of $\beta$ leads to a smooth profile while a large one leads to a sharp jump-like distribution. 
The unknown $\tilde{x}_{i}$, which represents the location of the jump center, is computed from constraint condition $\displaystyle \bar{q}_{i}^{T} = \frac{1}{\Delta x} \int_{x_{i-1/2}}^{x_{i+1/2}} \tilde{q}_{i}(x) \; dx$. 

Since the value given by hyperbolic tangent function $\tanh(x)$ lays in the region of $[-1,1]$, 
the value of THINC reconstruction function $\tilde{q}_{i}^{T}(x)$ is rigorously bounded by $\bar{q}_{i-1}$ and $\bar{q}_{i+1}$. 
Given the reconstruction function $\tilde{q}_{i}^{T}(x)$, we calculate the boundary values $q_{i+1/2}^{L,T}$ and $q_{i-1/2}^{R,T}$ by $q_{i+1/2}^{L,T}=\tilde{q}_{i}^{T}(x_{i+1/2})$ and $q_{i-1/2}^{R,T}=\tilde{q}_{i}^{T}(x_{i-1/2})$ respectively. 

In \cite{dengCF}, the effect of the sharpness parameter $\beta$ on numerical dissipation of the THINC scheme has been investigated with approximate dispersion relation (ADR) analysis. It is concluded that (i) with $\beta=1.1$, denoted by  $\beta_{s}$ hereafter, THINC has much smaller numerical dissipation than TVD scheme with Minmod limiter \cite{books}, and has similar but slightly better performance than the Van Leer limiter\cite{Van_Leer}; (ii) with a larger  $\beta$,  denoted by $\beta_{l}$, compressive or anti-diffusion effect will be introduced, which is preferred for discontinuous solutions. Based on this observation, a reconstruction strategy for both smooth and discontinuous solutions is  proposed by adaptively choosing the sharpness parameter $\beta$ with the BVD principle\cite{dengCF}. 

Following \cite{dengCF}, in this work we use THINC functions with different $\beta$ to realize non-oscillatory and less-dissipative reconstructions for various flow regions. A THINC reconstruction function $\tilde{q}_{i}^{Ts}(x)$  with $\beta_{s}=1.1$ is prepared to represent a relatively smoother solution, 
which gives the reconstructed values $q_{i+1/2}^{L,Ts}$ and $q_{i-1/2}^{R,Ts}$. 
To obtain better-resolved discontinuities, another THINC reconstruction function $\tilde{q}_{i}^{Tl}(x)$ with a larger $\beta_{l}=1.6$ is employed 
as an alternative candidate for reconstruction. The values at cell boundaries $q_{i+1/2}^{L,Tl}$ and $q_{i-1/2}^{R,Tl}$ are then obtained from $\tilde{q}_{i}^{Tl}(x)$ accordingly.

\subsubsection{The cascade BVD algorithm} 
As discussed above, the underlying fifth order upwind scheme can achieve optimal order for smooth region. However, numerical oscillations will appear in the presence of discontinuities such as contact discontinuities and shocks in compressible flow.  Limiting projection is the common practice used in existing methods to suppress numerical oscillation, where nonlinear weights are introduced to the polynomial function so as to degrade the interpolating polynomial to lower order. However, these limiting processes tend to  undermine the accuracy of the background linear high order schemes in spite of continuous efforts to explore better nonlinear weights. An ideal  limiting projection should maintain as much as possible the numerical properties of the scheme that uses the original polynomial reconstruction. In this work, we present a novel reconstruction scheme based on the BVD algorithm, where the linear high order schemes are directly used for smooth solutions. 

In \cite{dengCF}, a variant BVD algorithm was devised to minimize the total boundary variation (TBV), which implies that using monotonic interpolations in the neighboring cells results in smaller boundary variation values in presence of discontinuous solution, while high-order interpolations are preferred to minimize the boundary variations for smooth region. In the P4-THINC-BVD scheme presented in this work, reconstruction function is determined from the candidate interpolants with two-stage cascade BVD algorithm so as to minimize the TBV of the target cell. We denote the reconstruction function in the target cell ${\mathcal I}_{i}$ after the first stage BVD as $\tilde{q}_{i}^{<\text{I}>}(x)$ and the final reconstruction function after the second stage as $\tilde{q}_{i}^{<\text{II}>}(x)$.   

The two-stage cascade BVD algorithm is formulated as follows. 

\begin{description} 
\item{\bf (I) The first stage with polynomial of degree four and THINC($\beta_s$):} 
\begin{description} 
\item (I-I) Use the fifth order upwind scheme as the base reconstruction scheme and initialize the reconstructed function at the first stage as $\tilde{q}_{i}^{<\text{I}>}(x)=\tilde{q}_{i}^{P4}(x)$. 
\item (I-II) Calculate the TBV values for target cell ${\mathcal I}_{i}$ from the reconstruction of 4th degree polynomial as
\begin{equation}\label{Eq:TBVp4}
TBV_{i}^{P4}=\big|q_{i-1/2}^{L,P4}-q_{i-1/2}^{R,P4}\big|+\big|q_{i+1/2}^{L,P4}-q_{i+1/2}^{R,P4} \big|
\end{equation} 
and from the THINC function with a moderate steepness $\beta_{s}$ as    
\begin{equation}\label{Eq:TBVTs}
TBV_{i}^{Ts}=\big|q_{i-1/2}^{L,Ts}-q_{i-1/2}^{R,Ts}\big|+\big|q_{i+1/2}^{L,Ts}-q_{i+1/2}^{R,Ts} \big|.
\end{equation} 
\item (I-III) Modify the reconstruction function for cells $i-1$, $i$ and $i+1$ according to the following BVD algorithm
\begin{equation}\label{Eq:BVDlim-1}
\tilde{q}_{j}^{<\text{I}>}(x)=\tilde{q}_{j}^{Ts}(x), \ j=i-1,i,i+1;~~~{\rm if } \ \ TBV_{i}^{Ts}  < TBV_{i}^{P4}.
\end{equation} 
\end{description} 
\end{description} 

\begin{description} 
\item {\bf  (II) The second stage with $\tilde{q}_{i}^{<\text{I}>}(x)$  and THINC($\beta_l$):} 
\begin{description} 
\item (II-I) 
Compute the TBV using the reconstructed cell boundary values from stage (I) by     
\begin{equation}\label{Eq:TBVlim}
TBV_{i}^{<\text{I}>}=\big|q_{i-1/2}^{L,<\text{I}>}-q_{i-1/2}^{R,<\text{I}>}\big|+\big|q_{i+1/2}^{L,<\text{I}>}-q_{i+1/2}^{R,<\text{I}>} \big|, 
\end{equation}
and the TBV for THINC function of a larger steepness parameter  $\beta_{l}$ by 
\begin{equation}\label{Eq:TBVtl}
TBV_{i}^{Tl}=\big|q_{i-1/2}^{L,Tl}-q_{i-1/2}^{R,Tl}\big|+\big|q_{i+1/2}^{L,Tl}-q_{i+1/2}^{R,Tl} \big|. 
\end{equation}
\item (II-II)  Determine the final reconstruction function for cell ${\mathcal I}_{i}$ using the  BVD algorithm as
\begin{equation}
\tilde{q}_{i}^{<\text{II}>}(x)=\left\{
\begin{array}{l}
\tilde{q}_{i}^{Tl};~~~{\rm if } \ \ TBV_{i}^{Tl}  < TBV_{i}^{<\text{I}>}, \\
\tilde{q}_{i}^{<\text{I}>};~~~~\mathrm{otherwise}
\end{array}
\right..
\end{equation}

\item (II-III) Compute the reconstructed values on the left-side of $x_{i+\frac{1}{2}}$ and the right-side of $x_{i-\frac{1}{2}}$  respectively by 
\begin{equation}
q^{L}_{i+\frac{1}{2}}=\tilde{q}_{i}^{<\text{II}>}(x_{i+\frac{1}{2}}) \ \ {\rm and} \ \  q^{R}_{i-\frac{1}{2}}=\tilde{q}_{i}^{<\text{II}>}(x_{i-\frac{1}{2}}).
\end{equation}
 
\end{description} 
\end{description}

\begin{description}
\item {Remark 1. } 
The multi-stage cascade BVD algorithm enables to reinforce the desired numerical properties respectively at different stages. As formulated in the present method, the oscillation-free property is realized in stage (I), while stage (II) is devised to improve numerical dissipation. It implies a great flexibility in designing more sophisticated high-resolution schemes. 
\item {Remark 2. } 
Instead of the THINC function with $\beta_{s}$, other oscillation-suppressing  reconstruction schemes can also be used  in stage (I) to remove numerical oscillation.  We demonstrate other options of the candidate interpolants in appendix A, which shows that the BVD scheme with 4th degree polynomial and other TVD schemes can also work effectively to eliminate numerical oscillation.
\item {Remark 3. } 
The BVD scheme retrieves the spectral properties of the underlying  linear upwind scheme for all resolvable waves as shown later in 2.2.4 and appendix B. 
\item {Remark 4. }   
As demonstrated in \cite{dengCF} as well as the numerical results later in this paper, using the THINC function with a larger steepness parameter  ($\beta_{l}$) in stage (II) provides an efficient and robust scheme to reduce numerical dissipation.
 
\end{description}	

\subsubsection{Spectral property}
We analyzed the spectral property of the proposed scheme by using the approximate dispersion relation (ADR) method described in \cite{adr}. The numerical dissipation and dispersion of a scheme can be evaluated respectively from the imaginary and real  parts of the modified wavenumber. 

\begin{figure} [h]
	\begin{center}
		\subfigure{\centering\includegraphics[scale=0.35,trim={0.0cm 0.0cm 0.0cm 0.0cm},clip]{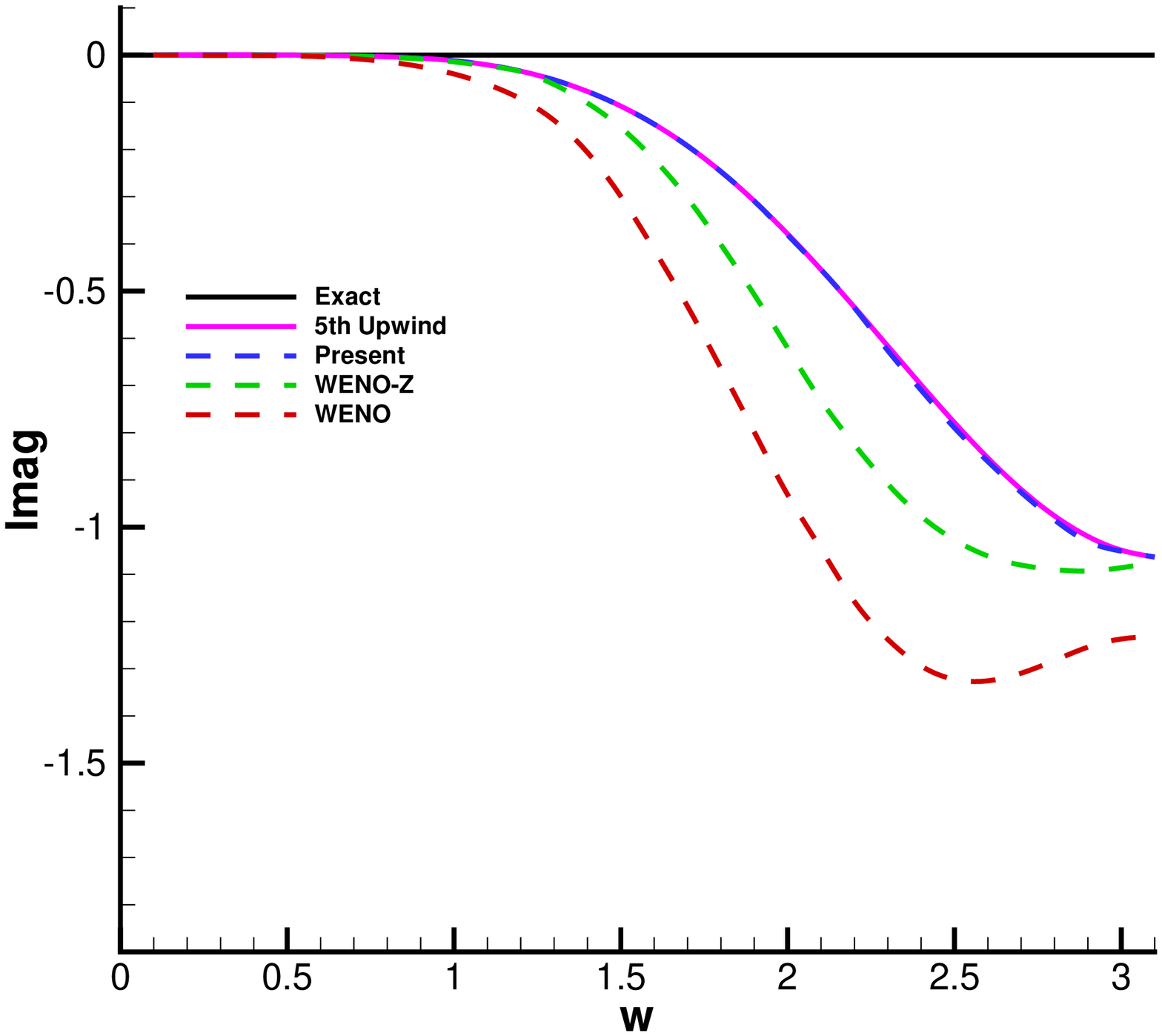}}
		\subfigure{\centering\includegraphics[scale=0.35,trim={0.0cm 0.0cm 0.0cm 0.0cm},clip]{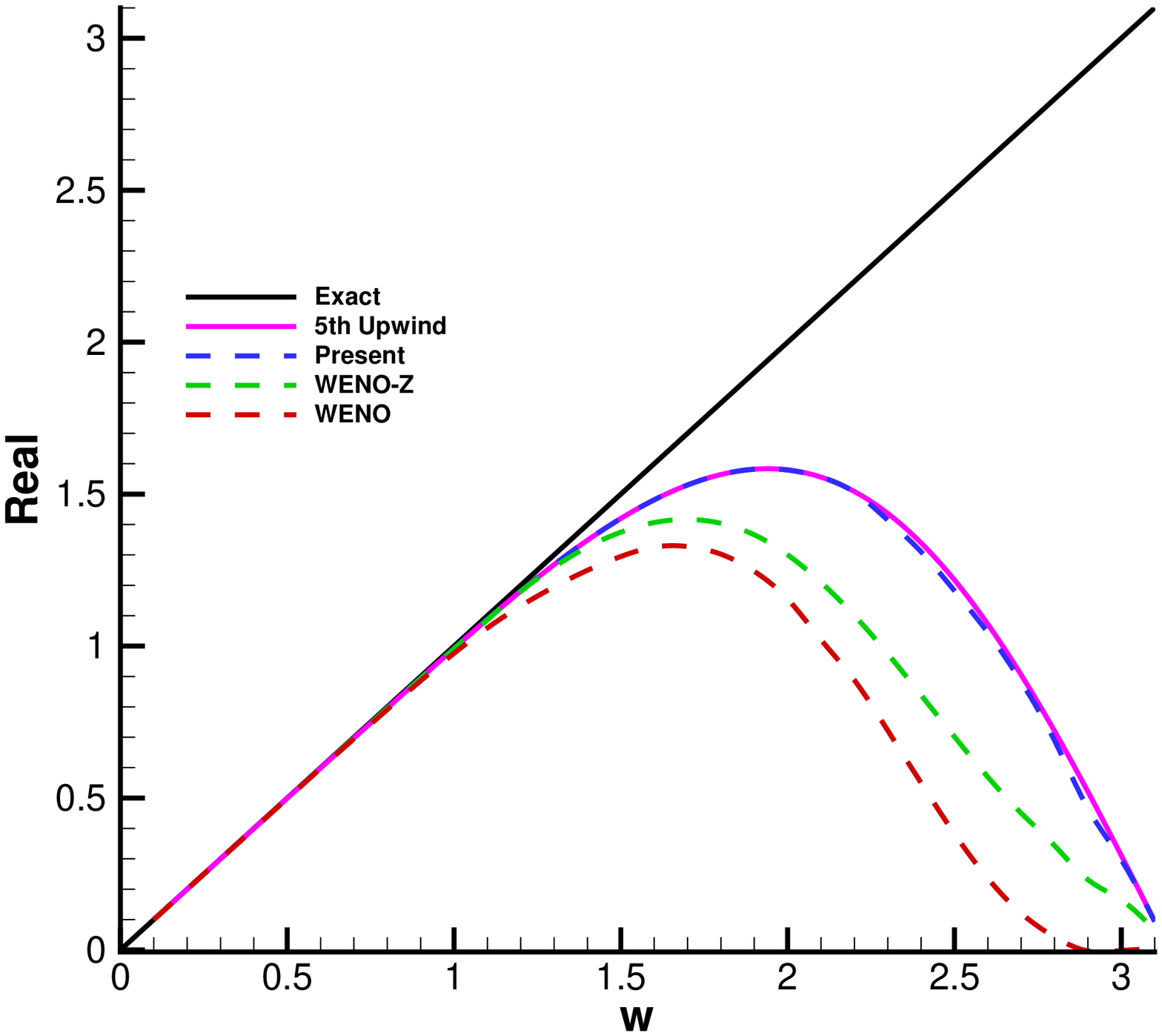}}
		\protect\caption{Approximate dispersion and dissipation properties for different schemes. Imaginary parts of modified wavenumber are shown in the left panel, while real parts are shown in the right.
			\label{fig:ADR}}
	\end{center}	
\end{figure}

The spectral property of the proposed scheme are shown in Fig.~\ref{fig:ADR} in which we also include the 5th order WENO scheme, WENO-Z scheme and linear 5th order upwind scheme using \eqref{Eq:5thuw} for comparisons. The  WENO schemes show significant discrepancies with the linear scheme for high wavenumber regime. It is partly owed to the fact that the WENO smoothness indicators tend to mis-interpret among high-wavenumber structures and discontinuities. Although the efforts in designing more sophisticated smoothness indicators, including WENO-Z \cite{wenoz} and more recent works \cite{TENO}, resolving  high-wavenumber  structures still remains challenging for WENO schemes.

On the contrary, the proposed scheme has almost the same spectral property as its underlying linear scheme. In appendix B, we  extensively examine the spectral properties of the BVD algorithm with other options of the candidate interpolants instead of the THINC function. It is found that the BVD algorithm using 4th degree polynomial and other existing TVD schemes can also retrieve the spectral properties of the 5th-order linear upwind scheme for all resolvable waves except the TVD scheme with the superbee limiter \cite{sweby84} which modulates the solution in the high-wavenumber region.    

The BVD method proposed in this work preserves the spectral property of the high order linear interpolation for all wave numbers, which surpasses all existing schemes using WENO methodology.   The present BVD method creates a new path to devise schemes that eliminate numerical oscillation yet enable to take the full advantage of the underlying line scheme of high-order polynomials. 
 
\section{Numerical results \label{sec:results}}

In this section, we show numerical results for benchmark tests of linear advection equation and Euler equations to verify the proposed scheme. All numerical results in one- and two- dimensions are compared with those of the WENO schemes. The CFL number $0.4$ is used in our tests unless specifically noted.

\subsection{Accuracy test for advection of one-dimensional sine wave}
This test is carried out on gradually refined grids to evaluate the convergence rate of the proposed scheme. The initial smooth distribution is given by
\begin{equation}
q\left(x\right)=\sin\left(\pi x\right), \ x\in\left[-1,1\right].
\end{equation}

We ran the computation for one period (at $t=2.0$) and summarize the numerical errors and the convergence rates for WENO, WENO-Z and the proposed scheme in Table \ref{Tab:rate}. Compared with WENO scheme and WENO-Z scheme, the proposed scheme generally has smaller $L_{1}$ and $L_{\infty}$ errors. Moreover, the $L_{1}$ and $L_{\infty}$ errors from the proposed scheme are exactly the same as those calculated by the 5th order linear upwind scheme, which is in line with the conclusion drawn from the spectral property analysis in the previous section. The superior accuracy of the proposed scheme is more remarkable when the mesh is coarse. Since the non-linear weights of the WENO schemes cannot recover their optimal counterparts under the coarse mesh, the accuracy degrades in spite of smooth initial profile. The WENO-Z scheme improves the behavior of non-linear weights by redesigning the smoothness indicator, which makes WENO-Z more accurate than the original WENO scheme even though it still does not recover exactly the 5th order linear upwind scheme. As stated in \cite{wenozn}, high gradients and fine smooth structures may be both regarded as discontinuities by the WENO smoothness indicators when grids are relatively coarse. On the contrary, the proposed scheme, without the non-linear weights, chooses the 5th order linear upwind scheme exactly following the BVD principle. It is revealed that the proposed scheme is able to achieve the highest possible accuracy of the linear high-order reconstruction for smooth solutions on all grid resolutions tested.    

\begin{table}[]
	\centering
	\caption{Numerical errors and convergence rates for linear advection test.  Comparisons are made among different schemes.}
	\label{Tab:rate}
	\begin{tabular}{l|lllll}
		\hline
		Schemes                                      & Mesh & $L_{1}$ errors & $L_{1}$ order & $L_{\infty}$ errors & $L_{\infty}$ order \\ \hline

				\multirow{5}{*}{WENO}
& 20   & $1.431\times10^{-3}$ &  & $2.511\times10^{-3}$ & \\		                     
& 40   & $4.473\times10^{-5}$ & 5.00 & $8.799\times10^{-5}$ & 4.83\\
& 80   & $1.396\times10^{-6}$ & 5.00 & $2.822\times10^{-6}$ & 4.96 \\
& 160  & $4.361\times10^{-8}$ & 5.00 & $8.487\times10^{-8}$ & 5.06 \\
& 320  & $1.361\times10^{-9}$ & 5.00 & $2.544\times10^{-9}$ & 5.06 \\ \hline
				\multirow{5}{*}{WENO-Z}
& 20   & $2.143\times10^{-4}$ &  & $3.537\times10^{-4}$ & \\		                     
& 40   & $6.399\times10^{-6}$ & 5.07 & $1.029\times10^{-5}$ & 5.10\\
& 80   & $2.004\times10^{-7}$ & 5.00 & $3.181\times10^{-7}$ & 5.02 \\
& 160  & $6.320\times10^{-9}$ & 4.99 & $9.955\times10^{-9}$ & 5.00 \\
& 320  & $2.035\times10^{-10}$ & 4.96 & $3.197\times10^{-10}$ & 4.96 \\ \hline
				\multirow{5}{*}{Present}
& 20   & $1.993\times10^{-4}$ &  & $3.154\times10^{-4}$ & \\		                     
& 40   & $6.372\times10^{-6}$ & 4.97 & $1.001\times10^{-5}$ & 4.98\\
& 80   & $2.003\times10^{-7}$ & 5.00 & $3.147\times10^{-7}$ & 4.99 \\
& 160  & $6.319\times10^{-9}$ & 4.99 & $9.926\times10^{-9}$ & 4.99 \\
& 320  & $2.035\times10^{-10}$ & 4.96 & $3.197\times10^{-10}$ & 4.96 \\ \hline
				\multirow{5}{*}{5th Upwind}
& 20   & $1.993\times10^{-4}$ &  & $3.154\times10^{-4}$ & \\		                     
& 40   & $6.372\times10^{-6}$ & 4.97 & $1.001\times10^{-5}$ & 4.98\\
& 80   & $2.003\times10^{-7}$ & 5.00 & $3.147\times10^{-7}$ & 4.99 \\
& 160  & $6.319\times10^{-9}$ & 4.99 & $9.926\times10^{-9}$ & 4.99 \\
& 320  & $2.035\times10^{-10}$ & 4.96 & $3.197\times10^{-10}$ & 4.96 \\ \hline
	\end{tabular}
\end{table}

\subsection{Accuracy test for advection of a smooth profile containing critical points}
\begin{figure} [h]
	\begin{center}
		\subfigure{\centering\includegraphics[scale=0.35,trim={0.5cm 0.5cm 0.5cm 0.5cm},clip]{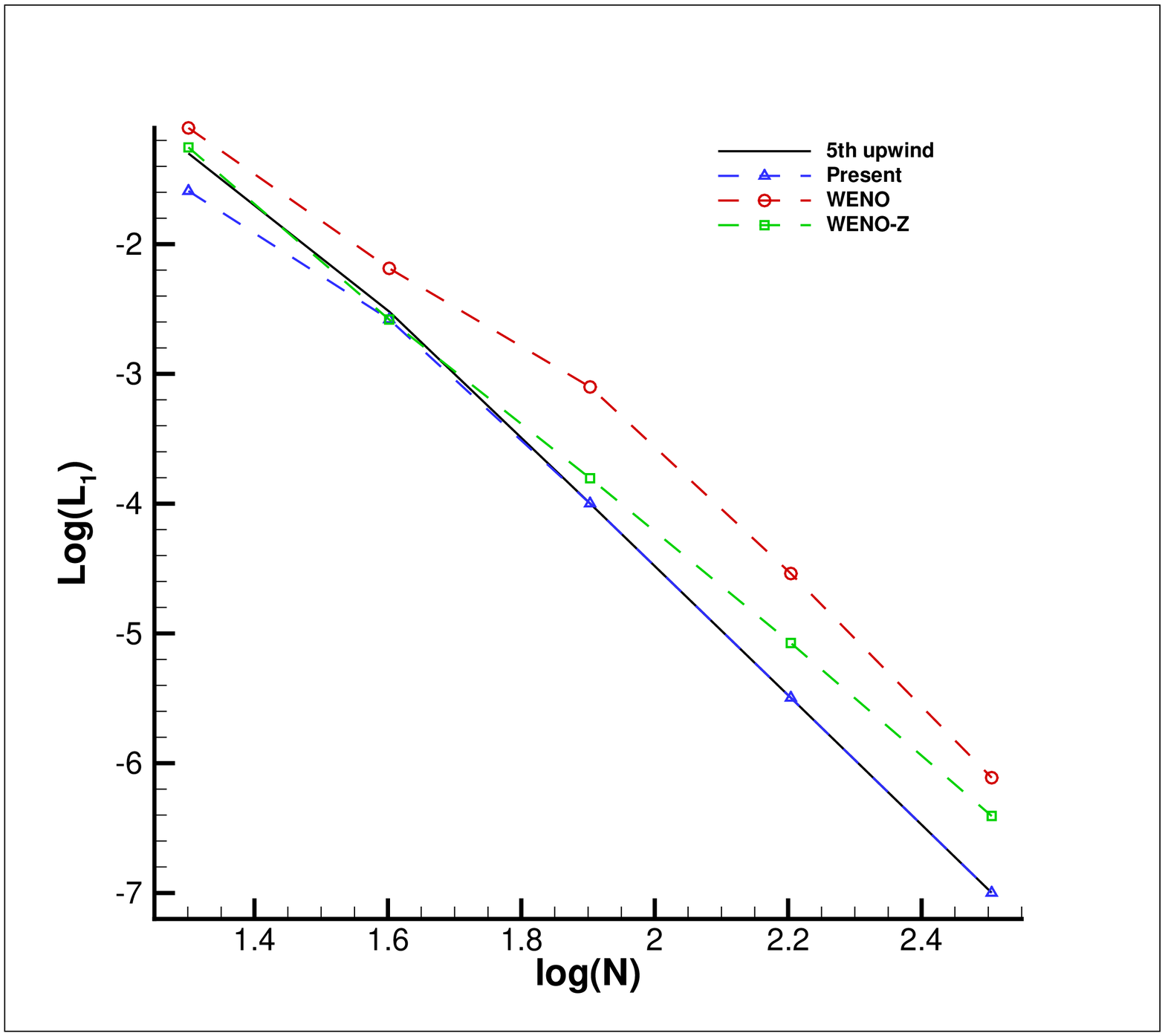}}
		\subfigure{\centering\includegraphics[scale=0.35,trim={0.5cm 0.5cm 0.5cm 0.5cm},clip]{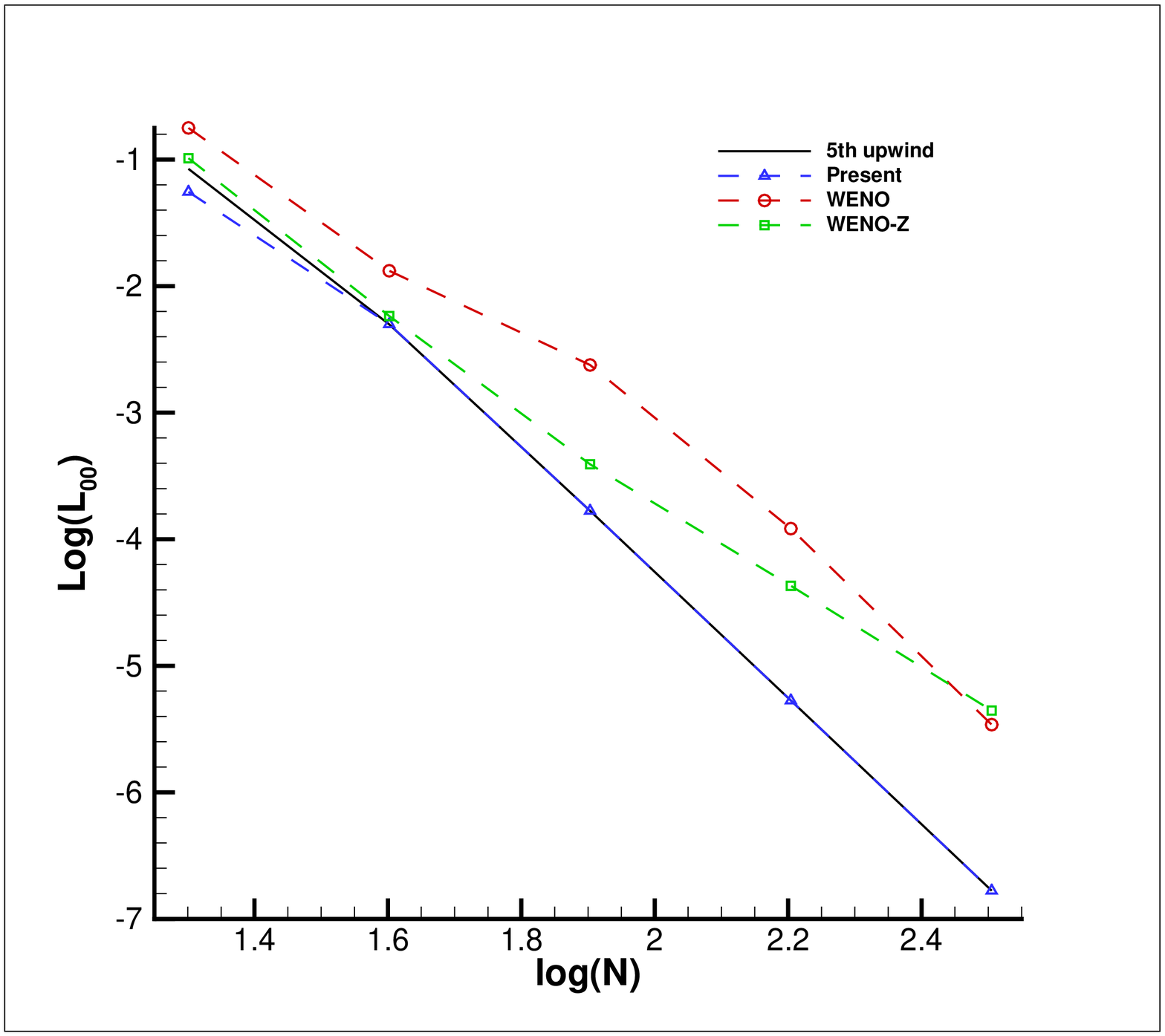}}
		\protect\caption{Numerical errors $L_{1}$ and $L_{\infty}$ as a function of the number of grid cells. Comparison are made between WENO, WENO-Z and the proposed P4-THINC-BVD scheme.
			\label{fig:sin4plot}}
	\end{center}	
\end{figure}

We conduct the accuracy test which is more challenging for numerical schemes to distinguish smooth and non-smooth profiles because the initial distribution contains critical points where high order derivative does not simultaneously vanish. It has been reported in \cite{WENOM} that WENO schemes do not reach their formal order of accuracy at critical points. Same as \cite{jiang96}, the initial condition is given by
\begin{equation}
q\left(x\right)=\sin^{4}\left(\pi x\right), \ x\in\left[-1,1\right].
\end{equation}
We ran the computation for one period (at $t=2.0$) and plot the numerical errors $L_{1}$ and $L_{\infty}$ for WENO, WENO-Z and the proposed scheme in Fig.~\ref{fig:sin4plot}. It can be seen that the proposed P4-THINC-BVD scheme exactly restores the accuracy of its underlying fifth order upwind scheme as the number of grid cells increase. However, both WENO and WENO-Z schemes lose accuracy even with refined mesh. The reason is that the mechanism of WENO non-linear adaptation will treat the critical point as non-smooth region. It is also noteworthy that at the coarsest mesh the P4-THINC-BVD even gives smaller errors than its underlying linear scheme. Since the distribution of initial profile is less smooth at the coarse mesh, the BVD algorithm will prefer THINC reconstruction which gives smaller numerical errors. This test shows that unlike WENO schemes the proposed P4-THINC-BVD can achieve the formal order of accuracy even for profiles containing critical points. Also, the BVD algorithm will choose the reconstruction function which gives smaller numerical dissipation errors.  

\subsection{Advection of complex waves} \label{sec:complex}
To examine the performance of the proposed scheme in solving profiles of different smoothness, we further simulated the propagation of a complex wave \cite{jiang96}, which includes both discontinuous and smooth solutions. This sample test is usually employed to verify the non-oscillatory property as well as numerical dissipation of shock-capturing schemes.

\begin{figure}[h]
	\includegraphics[scale=0.26,trim={0.9cm 0.9cm 0.9cm 0.9cm},clip]{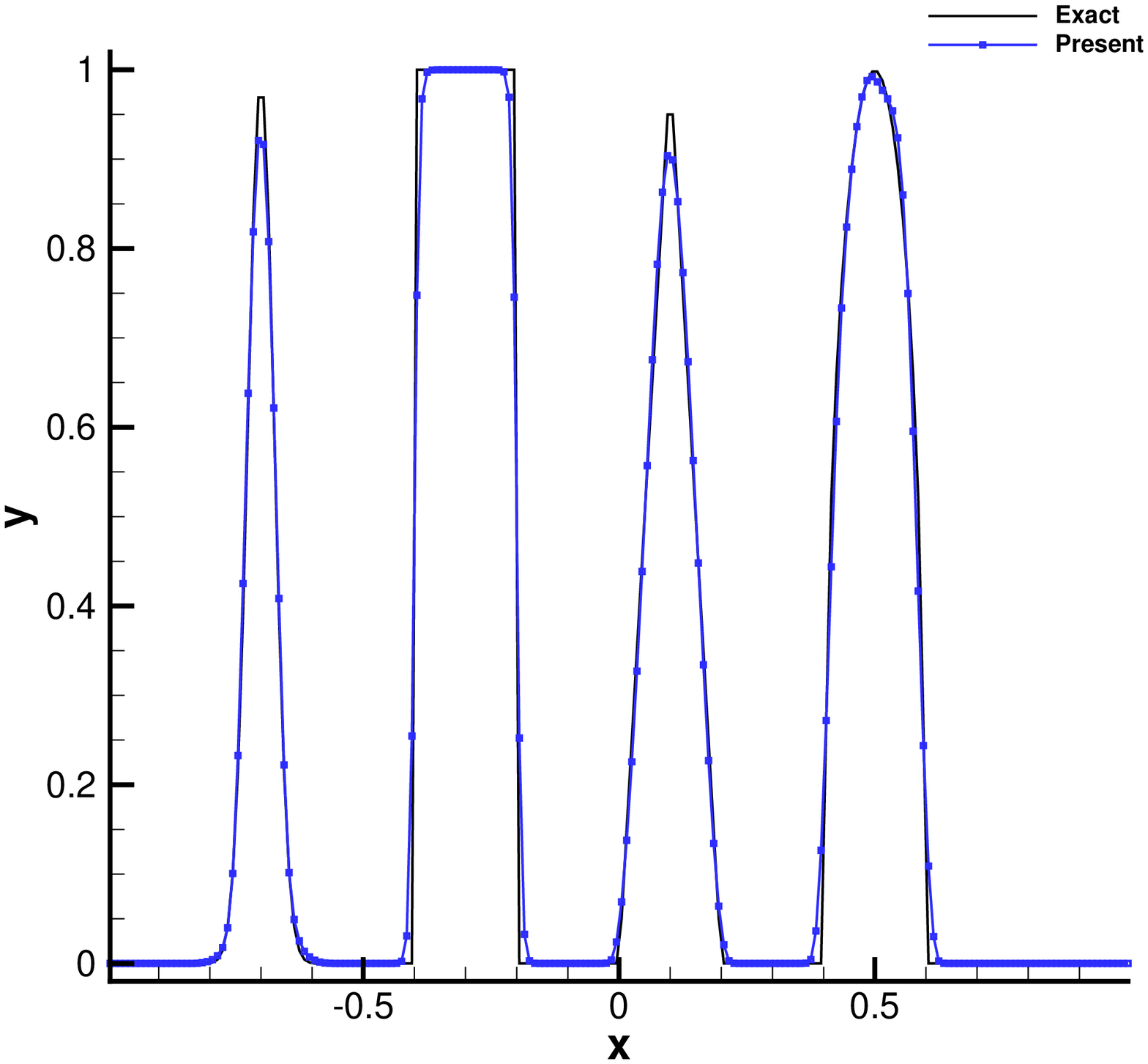}
	\includegraphics[scale=0.26,trim={0.9cm 0.9cm 0.9cm 0.9cm},clip]{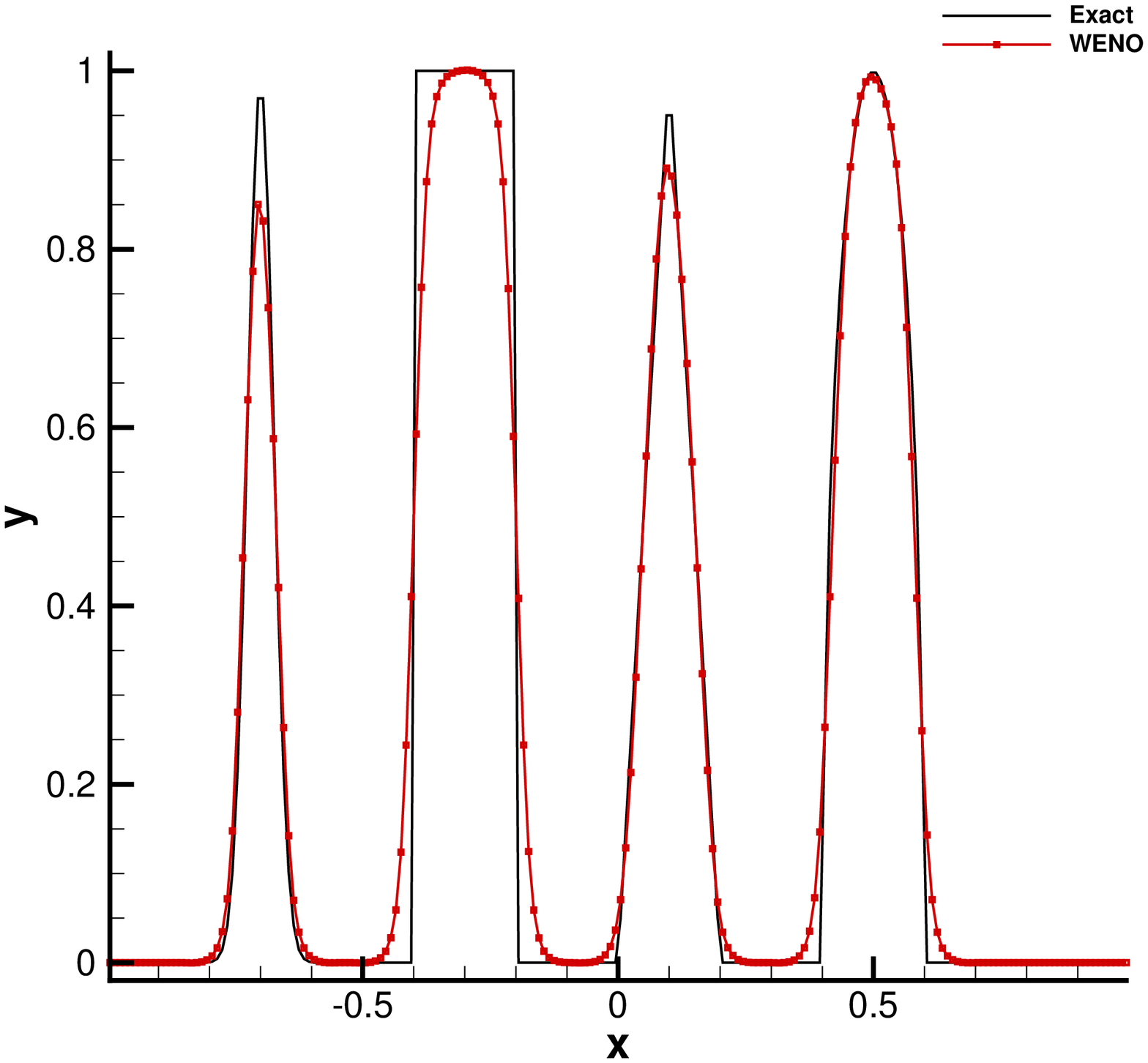} 
	\includegraphics[scale=0.26,trim={0.9cm 0.9cm 0.9cm 0.9cm},clip]{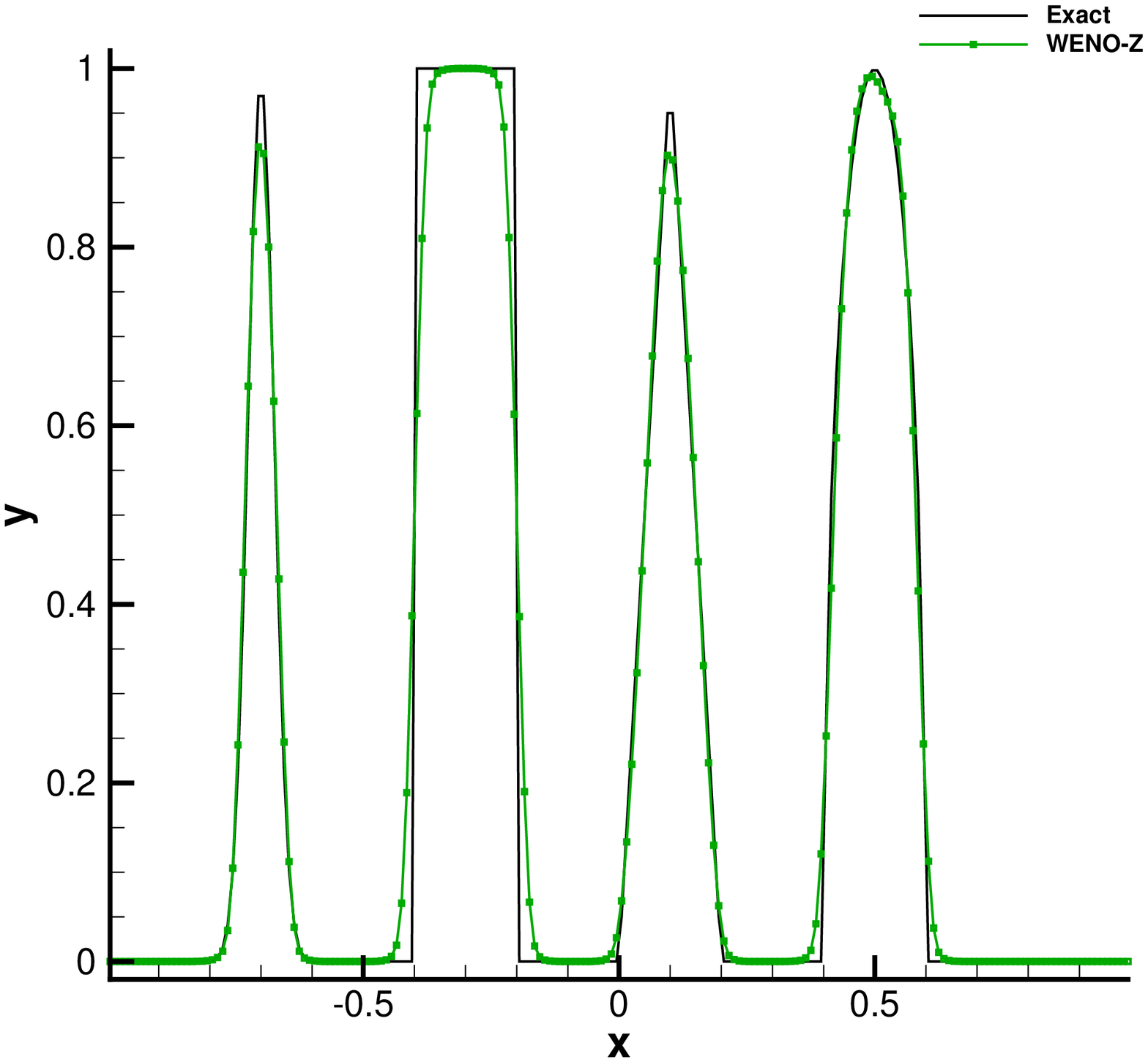} 
	\protect\caption{Numerical results for advection of complex waves. The numerical solutions at $t=2.0$ with 200 mesh cells are presented. Comparisons are made among the proposed scheme, the original WENO scheme and the WENO-Z scheme. 
		\label{fig:shujiang2}}	
\end{figure}

\begin{figure}[h]
	\includegraphics[scale=0.26,trim={0.9cm 0.9cm 0.9cm 0.9cm},clip]{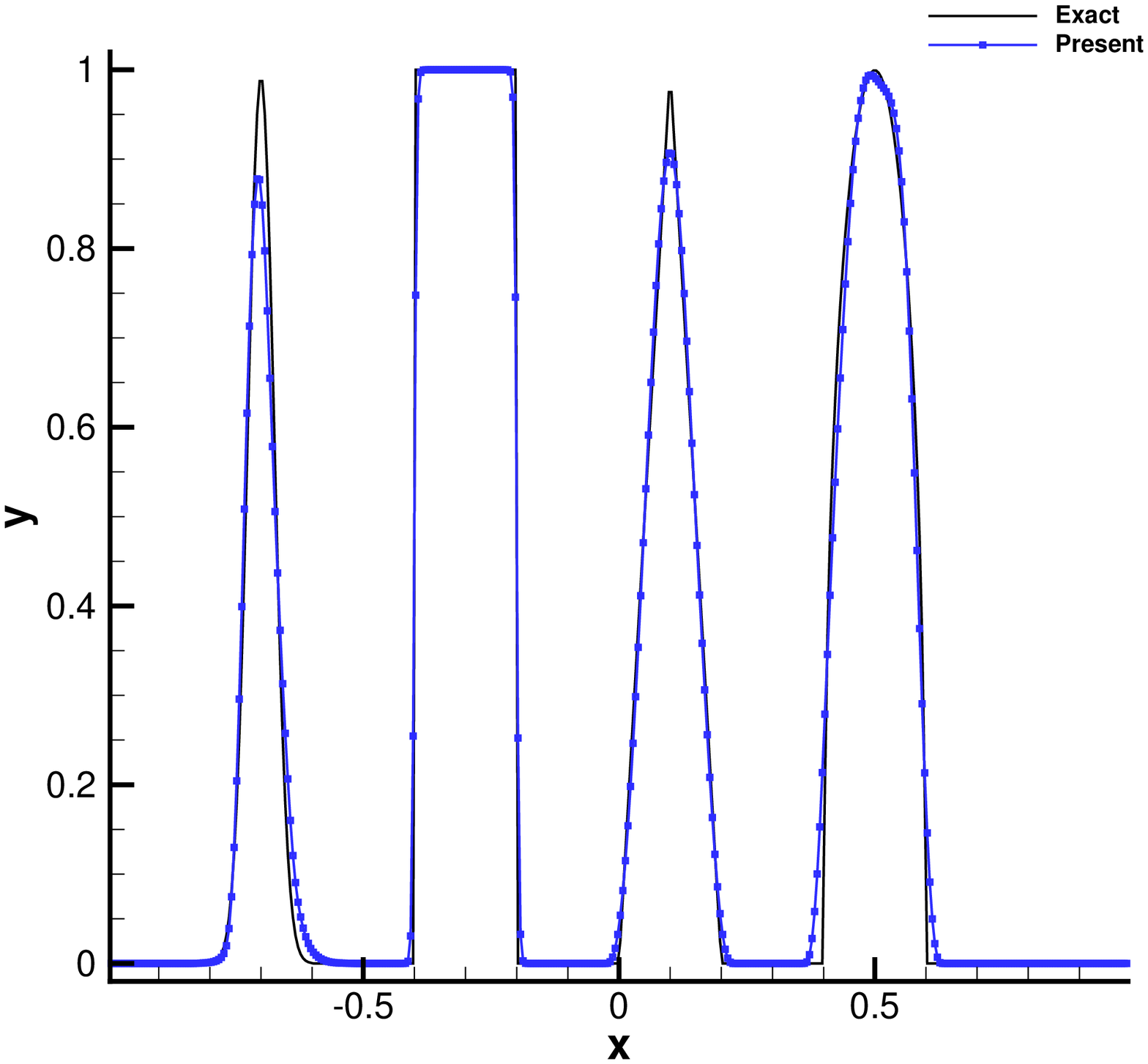}
	\includegraphics[scale=0.26,trim={0.9cm 0.9cm 0.9cm 0.9cm},clip]{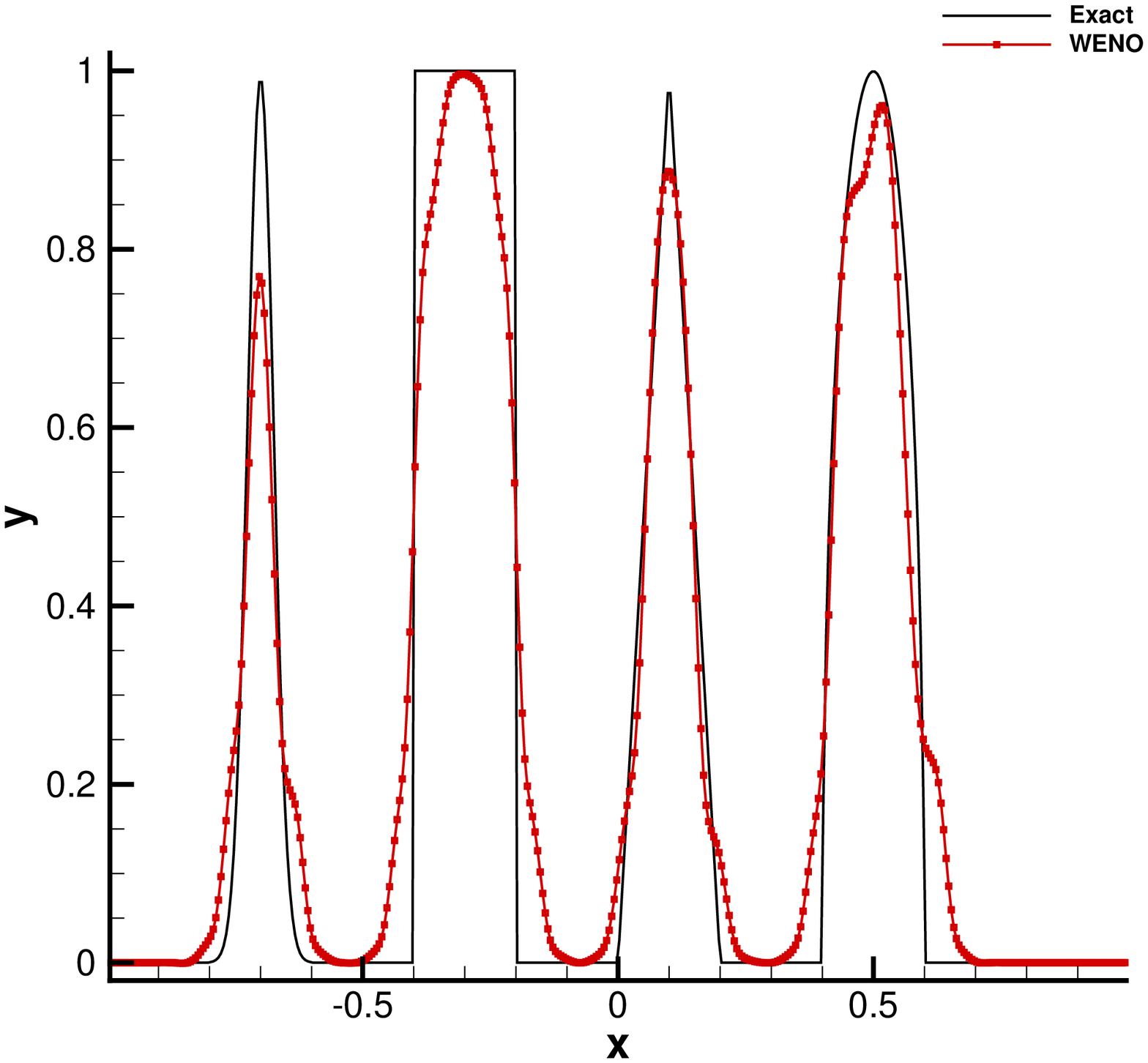} 
	\includegraphics[scale=0.26,trim={0.9cm 0.9cm 0.9cm 0.9cm},clip]{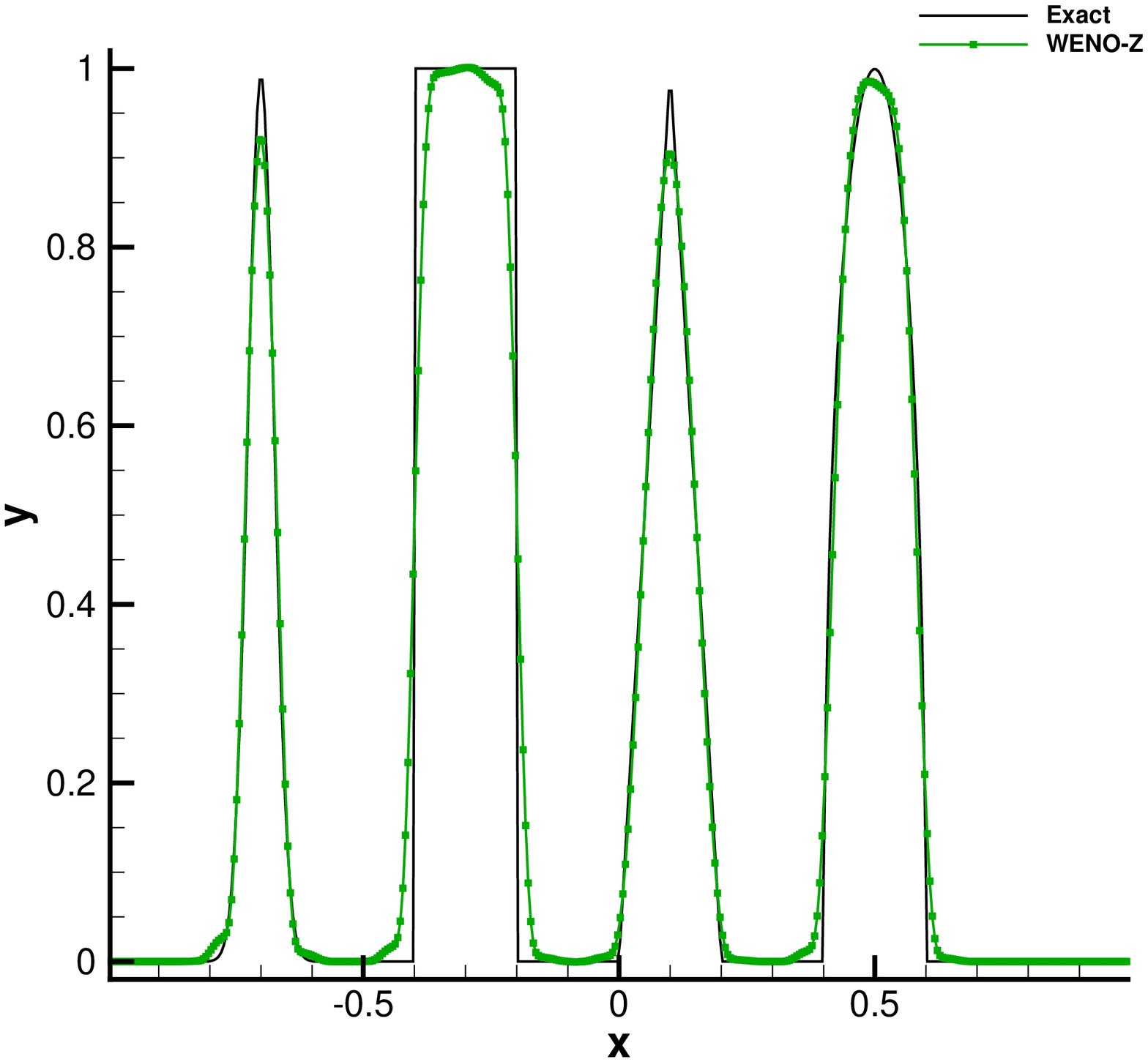} 
	\protect\caption{Same as Fig.~\ref{fig:shujiang2}, but with 400 mesh cells at $t=80$ which corresponds to $4.0\times10^{4}$ steps.
		\label{fig:shujiang80}}	
\end{figure}

\begin{figure}[h]
	\includegraphics[scale=0.26,trim={0.9cm 0.9cm 0.9cm 0.9cm},clip]{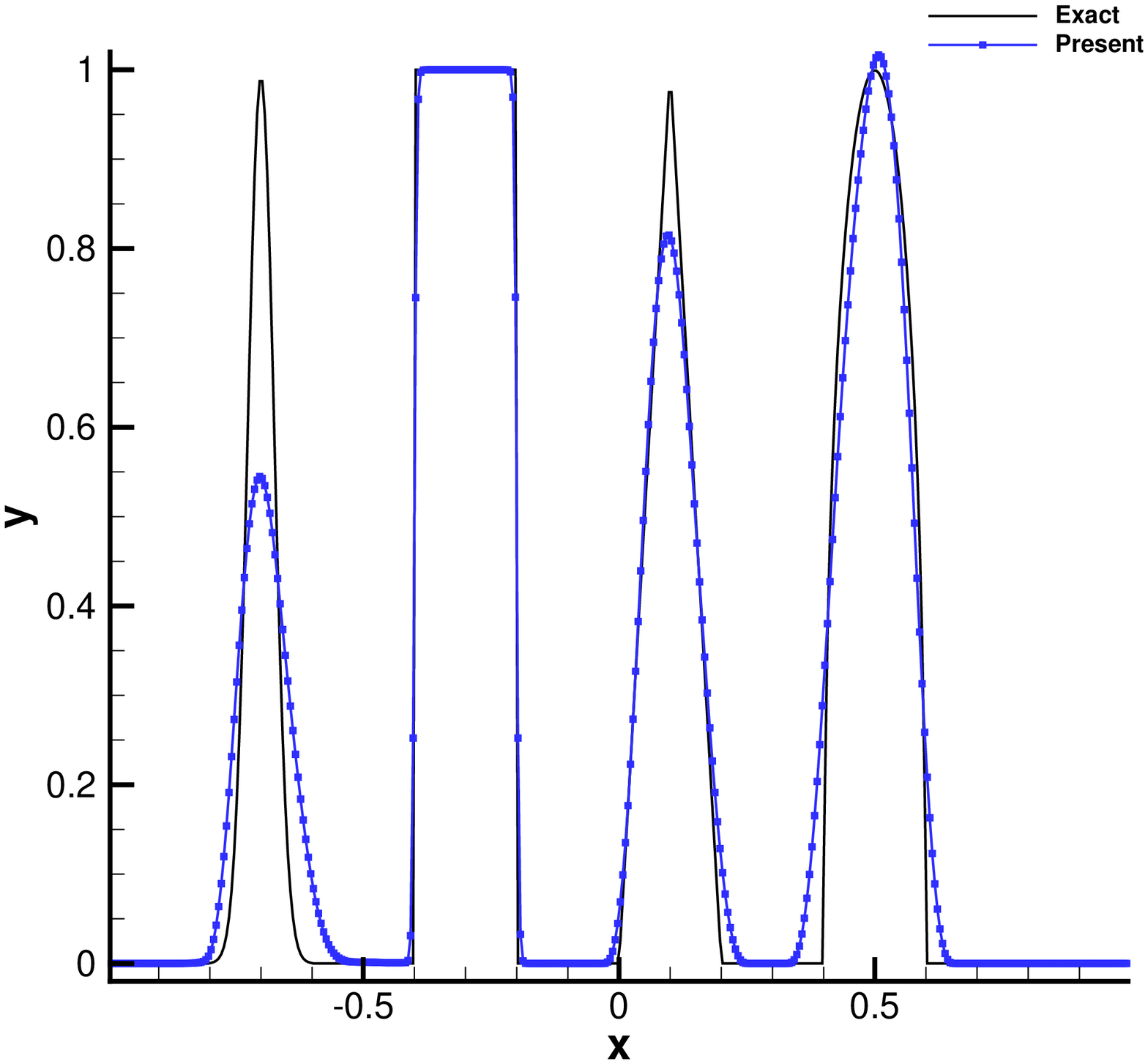}
	\includegraphics[scale=0.26,trim={0.9cm 0.9cm 0.9cm 0.9cm},clip]{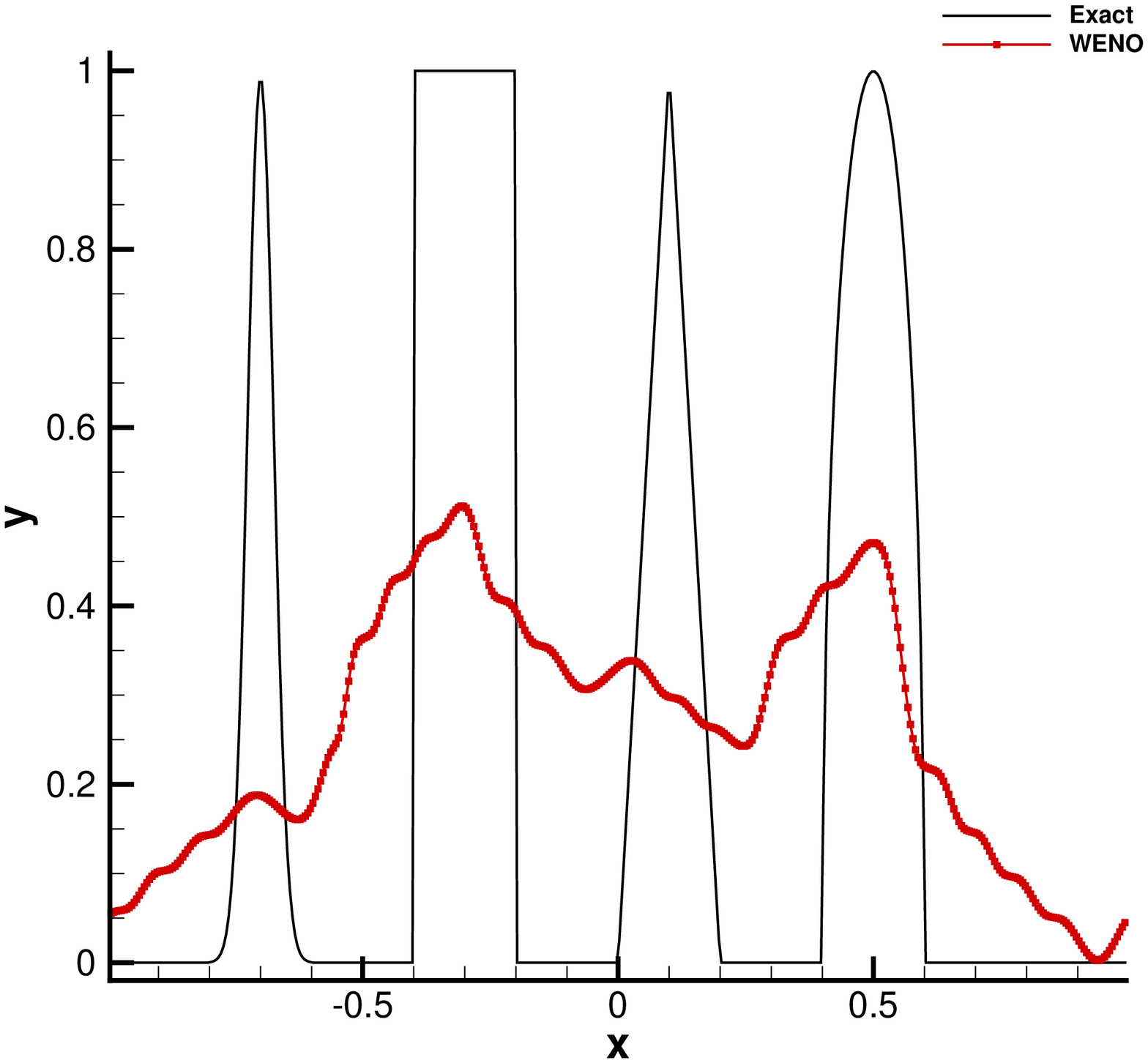} 
	\includegraphics[scale=0.26,trim={0.9cm 0.9cm 0.9cm 0.9cm},clip]{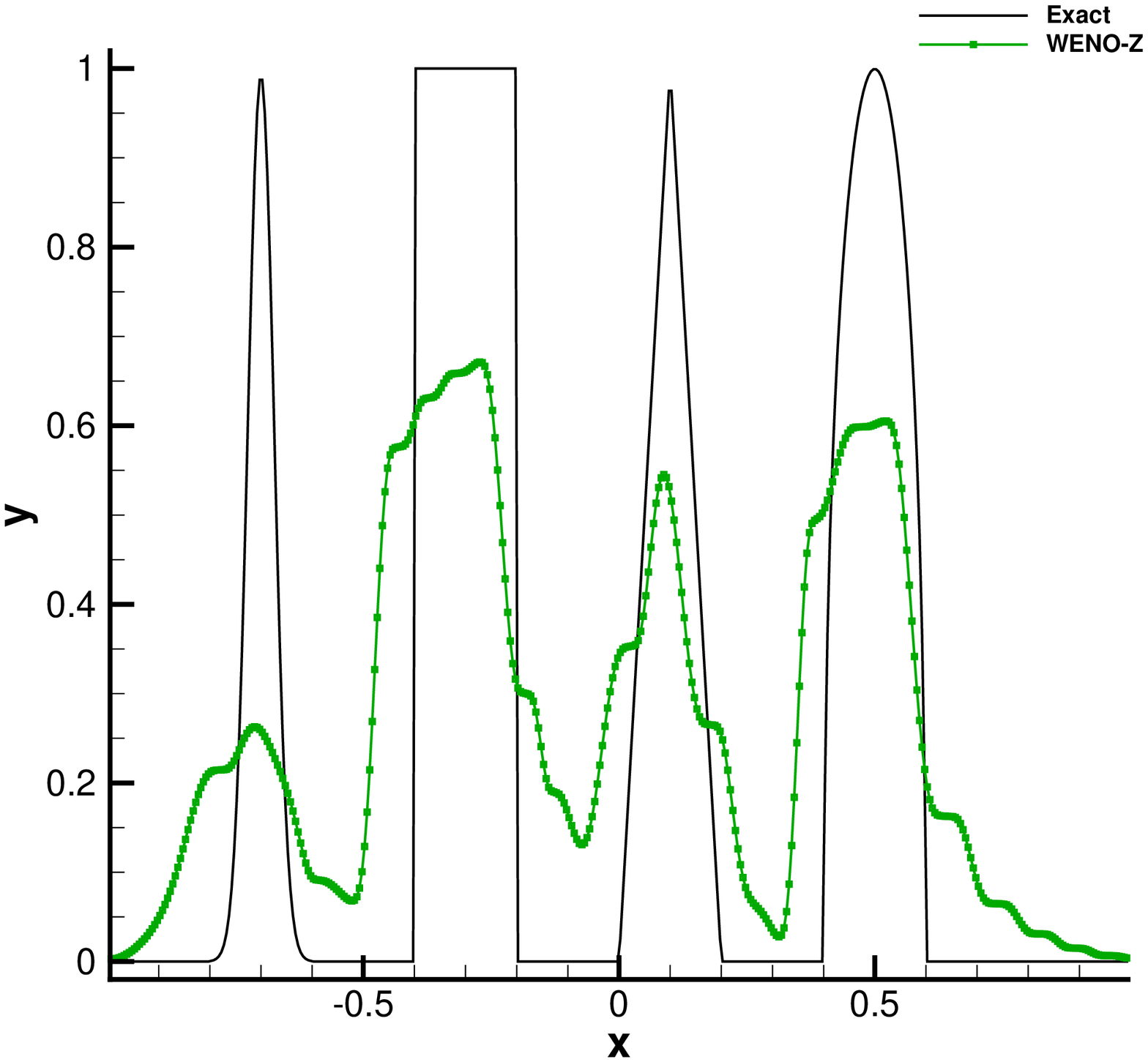} 
	\protect\caption{Same as the Fig.~\ref{fig:shujiang2} but with 400 mesh cells at $t=2000$ which corresponds to $1.0\times10^{6}$ steps.
		\label{fig:shujiang2000}}	
\end{figure}

As a commonly reported case in the literature, the numerical results at $t=2.0$ ($1.0\times10^{3}$ steps) on a 200-cell mesh of  the proposed scheme are plotted in Fig.~\ref{fig:shujiang2} against the original WENO and WENO-Z schemes. The proposed scheme effectively suppresses numerical oscillation as the WENO schemes, while resolving more accurately the discontinuous profile. 
In order to evaluate the performance of the schemes in simulations for long time periods, we show the numerical results on a 400-cell mesh at $t=80$ ($4.0\times10^{4}$ steps) and  $t=2000$ ($1.0\times10^{6}$ steps) in Figs.~\ref{fig:shujiang80} and \ref{fig:shujiang2000} respectively. More profoundly, the advantage of the proposed scheme becomes more visible in these cases.  From the results of  $t=80$ and $t=2000$ , it can be seen that the WENO schemes gradually distort the initial profile and continuously worsen the solution quality, while the proposed scheme adequately resolves both discontinuities and smooth solutions even after one million steps of computation as an extreme case. It reveals that the present scheme is well suited for simulating physical problems where the free modes are of importance but easily contaminated by numerical errors.    

\subsection{Sod's problem}
\begin{figure}
	\includegraphics[scale=0.35,trim={0.5cm 0.5cm 0.5cm 0.5cm},clip]{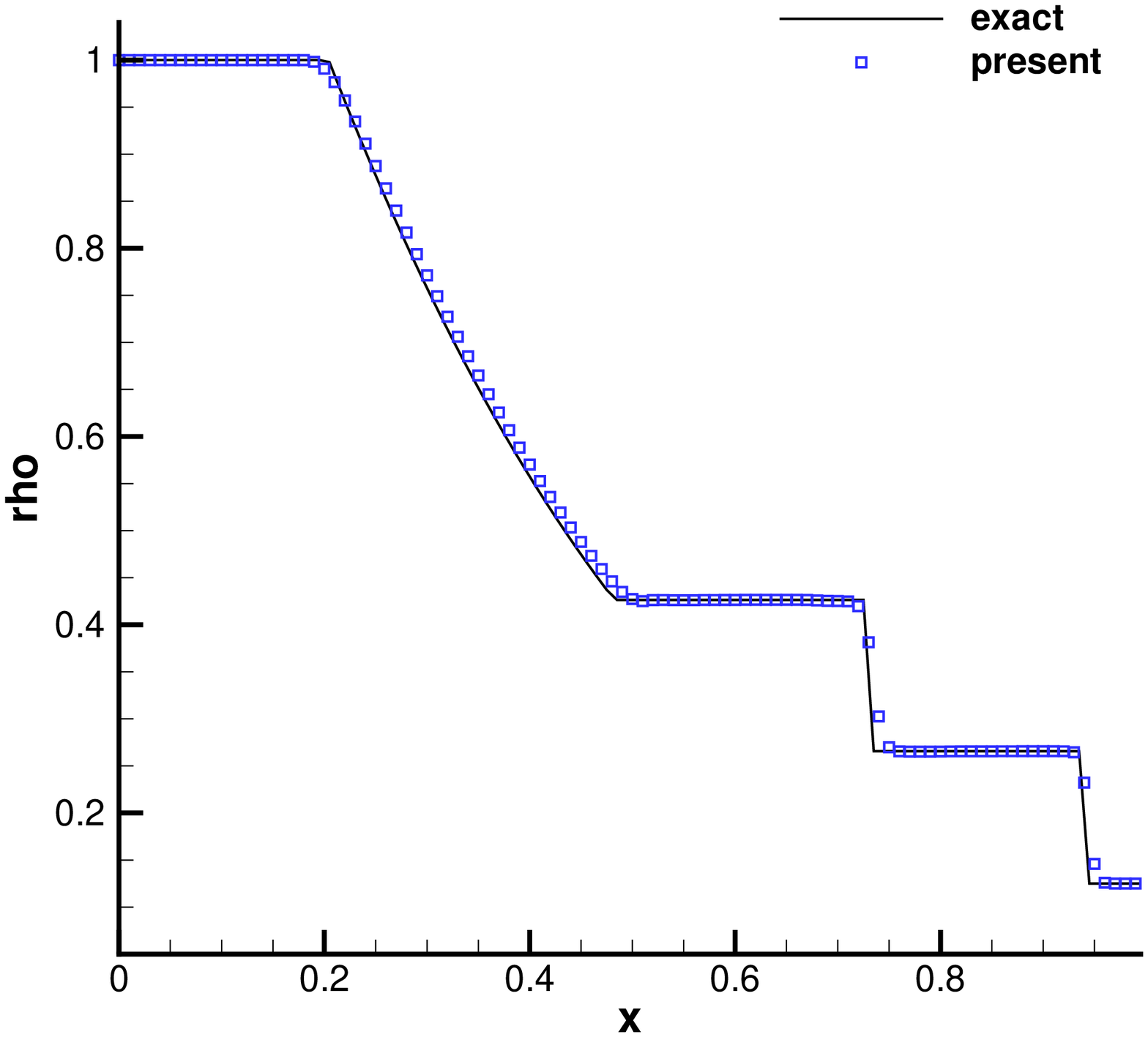}
	\includegraphics[scale=0.35,trim={0.5cm 0.5cm 0.5cm 0.5cm},clip]{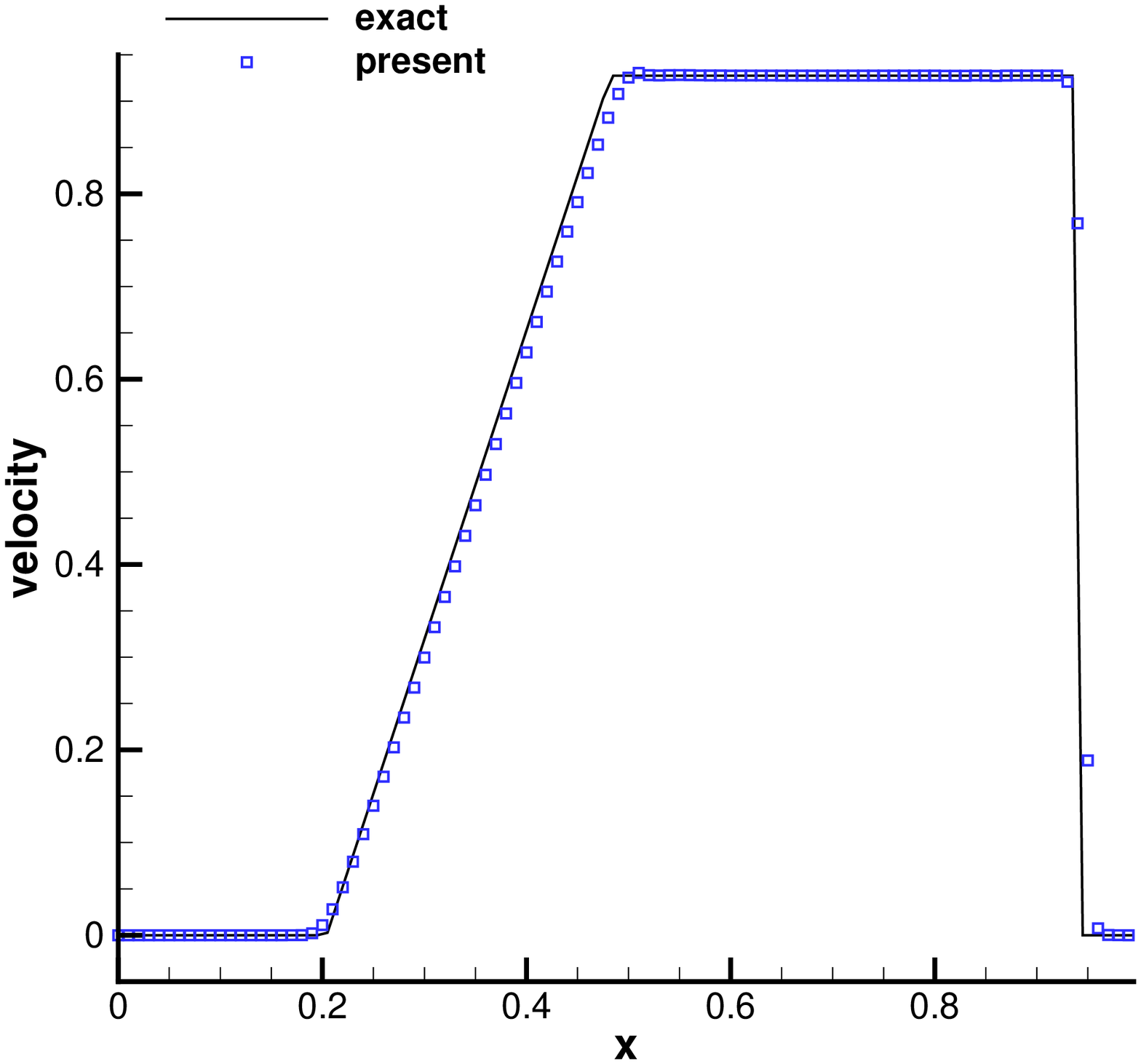} 
	\protect\caption{Numerical results of Sod's problem for density (left) and velocity (right)   fields at $t = 0.25$ with $100$ mesh cells. 
    \label{fig:sodrho}}	
\end{figure}

The Sod problem, which is one of famous benchmark tests for shock-capturing schemes, is employed here to test the performance of numerical schemes in solving Euler equations, which include discontinuous solutions like shock front and contact discontinuity. The initial distribution on computational domain $[0,1]$ is specified as \cite{sod}
\begin{equation}
\left(\rho_{0},\ u_{0},\ p_{0}\right)=\left\{
\begin{array}{ll}
\left(1,\ 0,\ 1\right), & 0 \leq x \leq 0.5 \\
\left(0.125,\ 0,\ 0.1\right), & \mathrm{otherwise}
\end{array}
\right..
\end{equation}
The computation is carried out on a mesh with 100 uniform cells up to $t=0.25$. The numerical results calculated from the proposed scheme are shown in Fig.~\ref{fig:sodrho} for density and velocity fields respectively. It is observed that the proposed scheme can solve both the shock wave and contact discontinuity without numerical oscillations. Moreover, different from other existing shock-capturing schemes which produce diffusive solution around contact discontinuities, the proposed scheme can resolve the density jump at the contact discontinuity with only two or three cells. 

\subsection{Lax's problem}
\begin{figure}
  {\centering\includegraphics[scale=0.35,trim={0.5cm 0.5cm 0.5cm 0.5cm},clip]{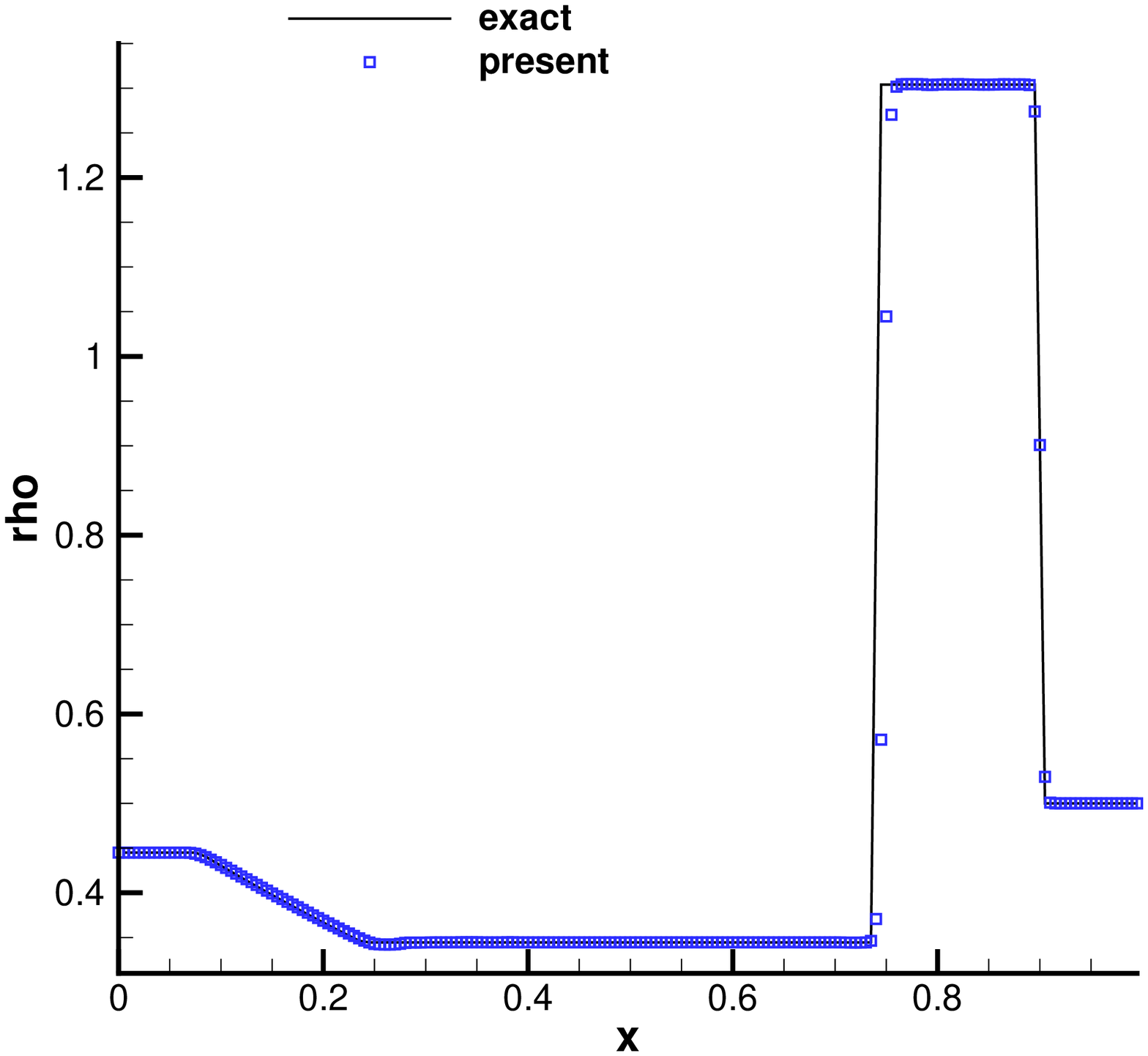}} 
  {\centering\includegraphics[scale=0.35,trim={0.5cm 0.5cm 0.5cm 0.5cm},clip]{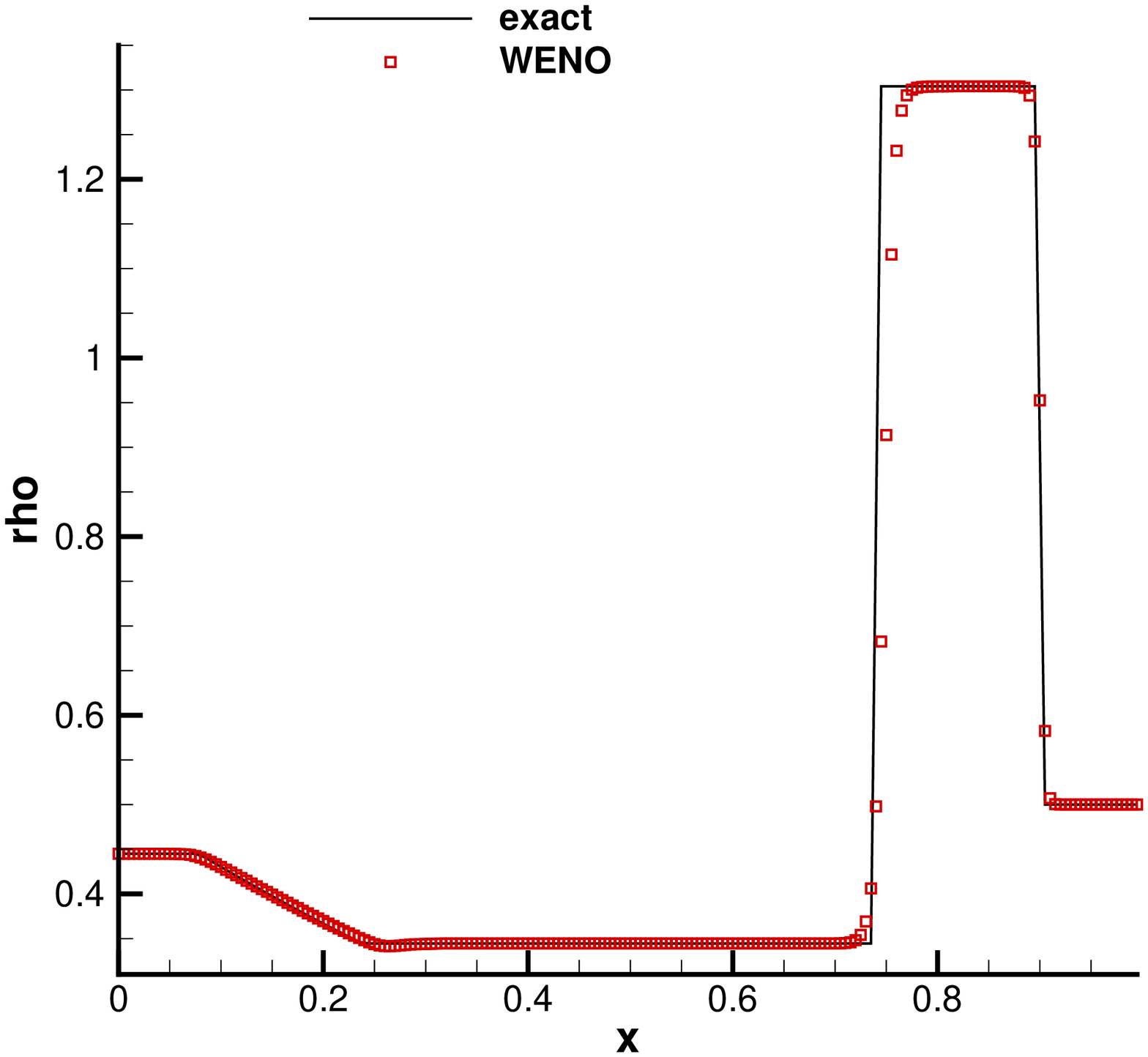}}
  \protect\caption{Numerical results of Lax's problem for density field at time $t = 0.16$ with $200$ mesh cells. Comparisons are made between the proposed scheme (left panel) and WENO scheme (right panel) against the exact solution (black straight line).
    \label{fig:lax}}	
\end{figure}

To evaluate the ability of the proposed numerical scheme 
to capture relatively strong shock, we solve the Lax problem \cite{shu_eno1} in this subsection. The initial condition is given by
\begin{equation}
\left(\rho_{0},\ u_{0},\ p_{0}\right)=\left\{
\begin{array}{ll}
\left(0.445,\ 0.698,\ 3.528\right), & 0 \leq x \leq0.5\\
\left(0.5,\ 0.0,\ 0.571\right), & \mathrm{otherwise}
\end{array}
\right..
\end{equation}
With 200 uniform cells, the numerical result at $t=0.16$ of  the proposed scheme is plotted in  Fig.~\ref{fig:lax} in comparison with the original WENO scheme.
The proposed scheme produces accurate solutions without numerical oscillations. Again, the contact discontinuity is resolved very sharply while the result of the WENO scheme looks more diffusive. Compared with other published works \cite{wenozn,TENO} which simulate this case with improved schemes using WENO methodology, the current result is still one of best.

\begin{figure}
	\begin{center}
    \subfigure[]
	{\centering\includegraphics[scale=0.35,trim={0.5cm 0.5cm 0.5cm 0.5cm},clip]{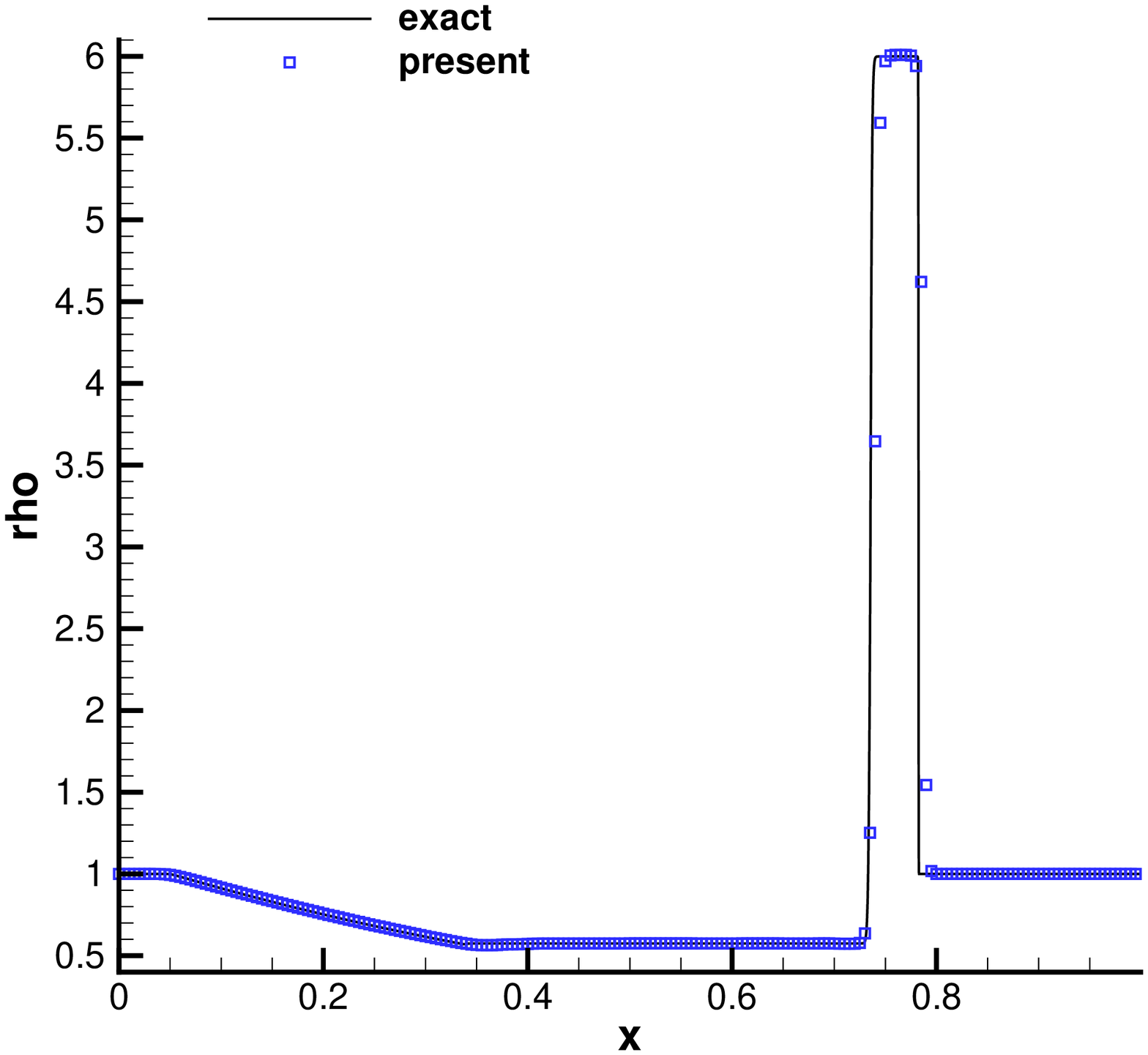}}
    \subfigure[]
	{\centering\includegraphics[scale=0.35,trim={0.5cm 0.5cm 0.5cm 0.5cm},clip]{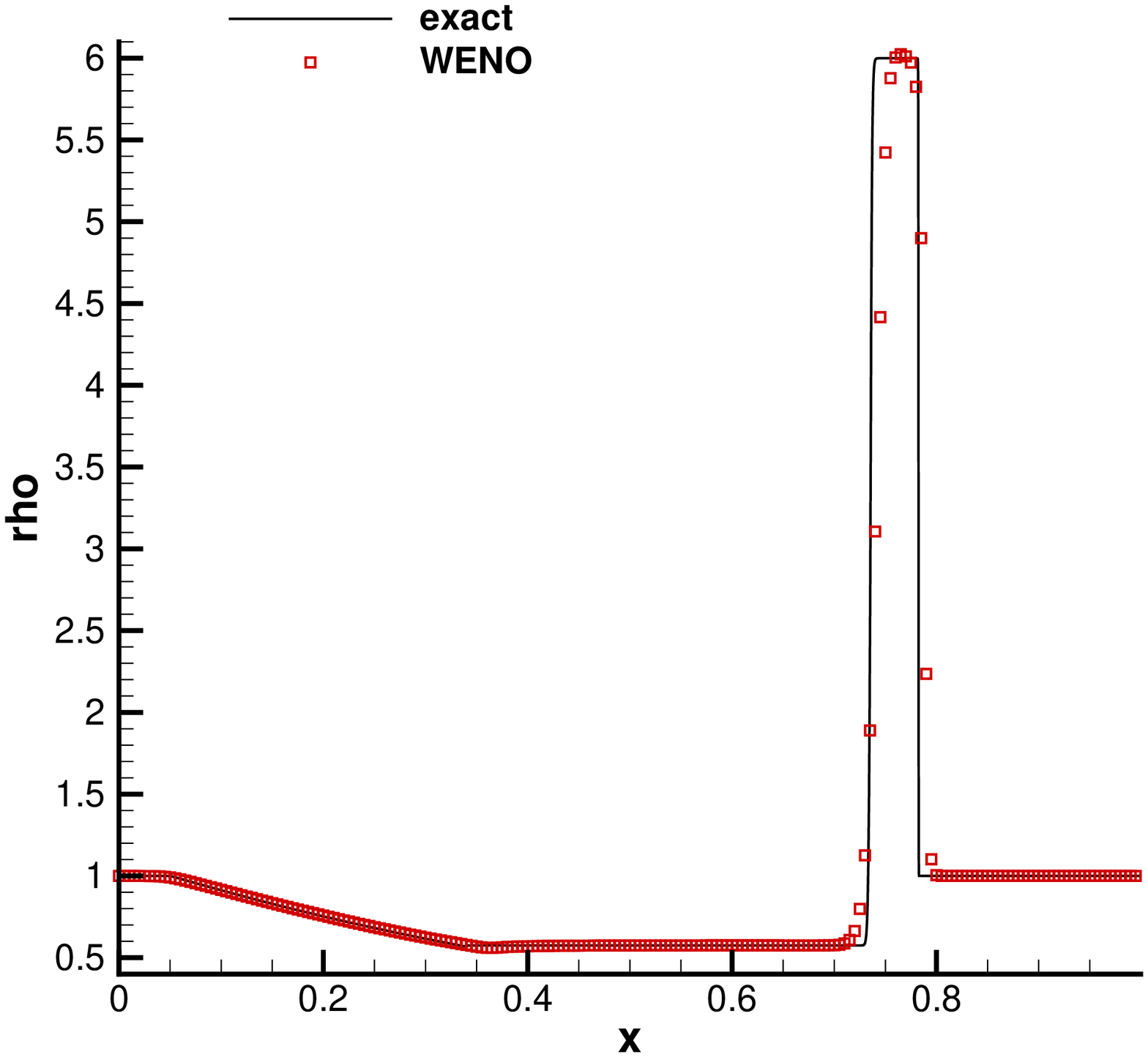}}
	\subfigure[]
	{\centering\includegraphics[scale=0.35,trim={0.5cm 0.5cm 0.5cm 0.5cm},clip]{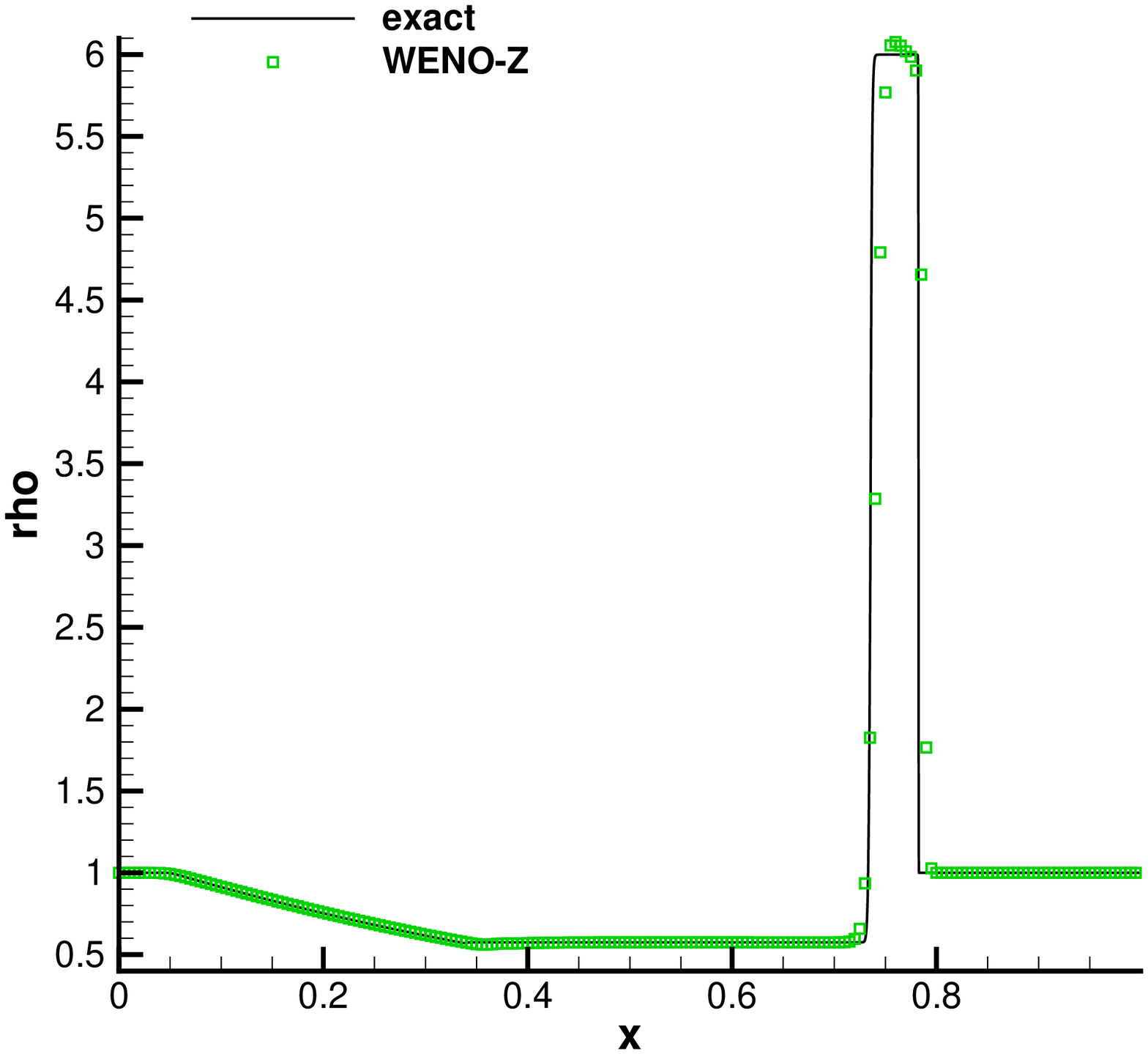}}
		\subfigure[]
	{\centering\includegraphics[scale=0.35,trim={0.5cm 0.5cm 0.5cm 0.5cm},clip]{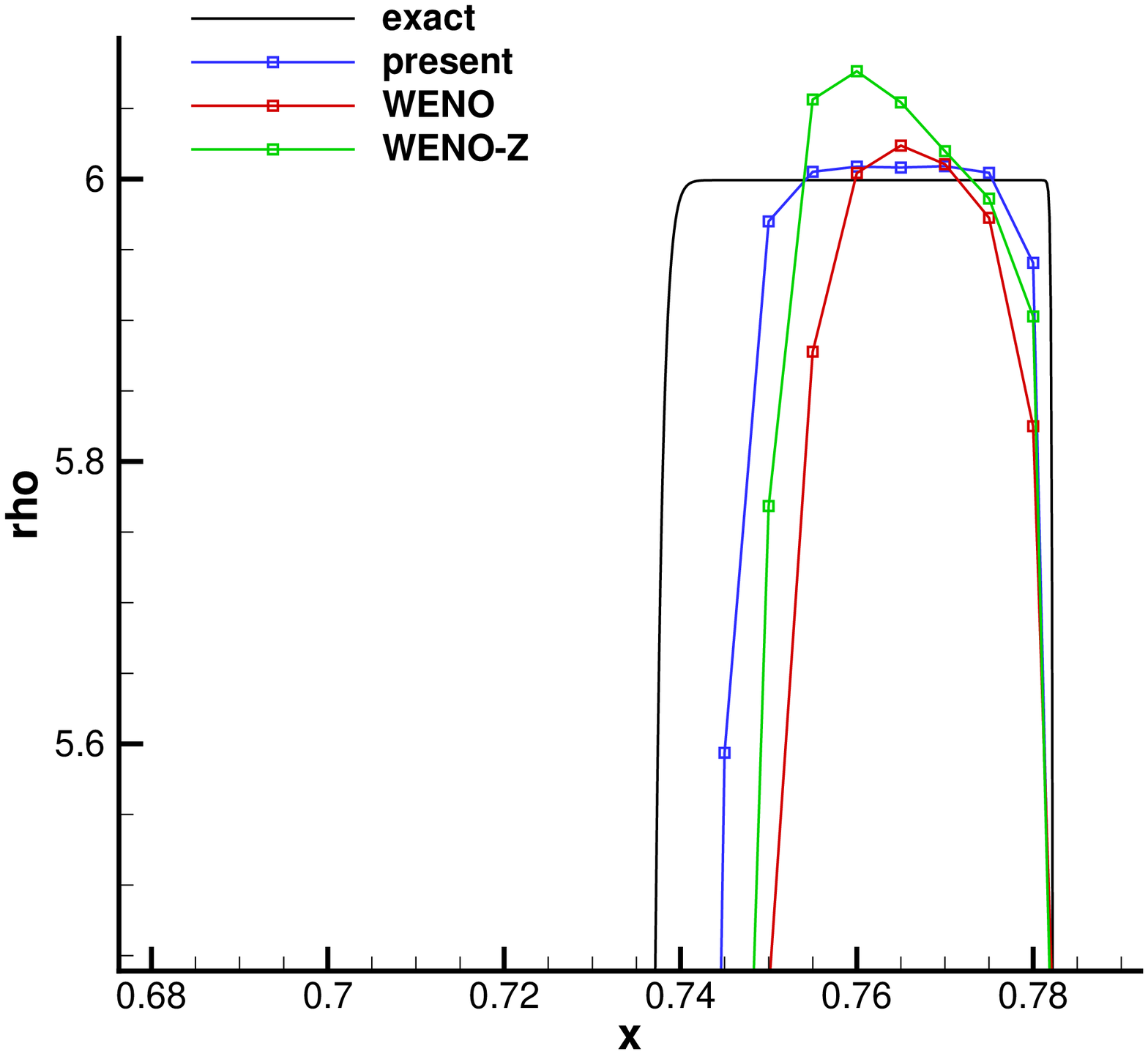}}
	\protect\caption{Numerical results of strong Lax's problem with Mach 198 for density field at time $t = 0.012$ with $200$ cells. Comparisons are made among different schemes. In the right bottom panel, the zoomed region around density peak is presented.
		\label{fig:stronglax}}	
	\end{center}
\end{figure}

To further test the performance of current scheme on problems with stronger discontinuities, we conduct another Lax problems with following initial conditions:
\begin{equation}
\left(\rho_{0},\ u_{0},\ p_{0}\right)=\left\{
\begin{array}{ll}
\left(1.0,\ 0.0,\ 1000.0\right), & 0 \leq x \leq0.5\\
\left(1.0,\ 0.0,\ 0.01\right), & \mathrm{otherwise}
\end{array}
\right..
\end{equation}
With initial high pressure ratio, a right-moving Mach 198 shock and a contact is generated. The computation lasts until time $t=0.012$ with 200 cell elements. The numerical solutions from different schemes for density field are plotted in Fig.~\ref{fig:stronglax}. In the same manner, the proposed scheme can resolve the contact discontinuity with superior accuracy in comparison with other schemes. For more detailed comparison, we show the zoomed region around the highly compressed region in Fig.~\ref{fig:stronglax}(d).  It can be seen that the proposed scheme gives more faithful solution with reduced errors in both dissipation and overshooting. The WENO-Z scheme produces the largest overshoot among the tested schemes as  WENO-Z scheme assigns larger weights to less-smooth stencils in order to reduce numerical dissipation. This test justifies the proposed scheme as a robust solver for strong discontinuities with high accuracy. 

\subsection{Shock/density wave interaction problem}
\begin{figure}
	{\centering\includegraphics[scale=0.35,trim={0.5cm 0.5cm 0.5cm 0.5cm},clip]{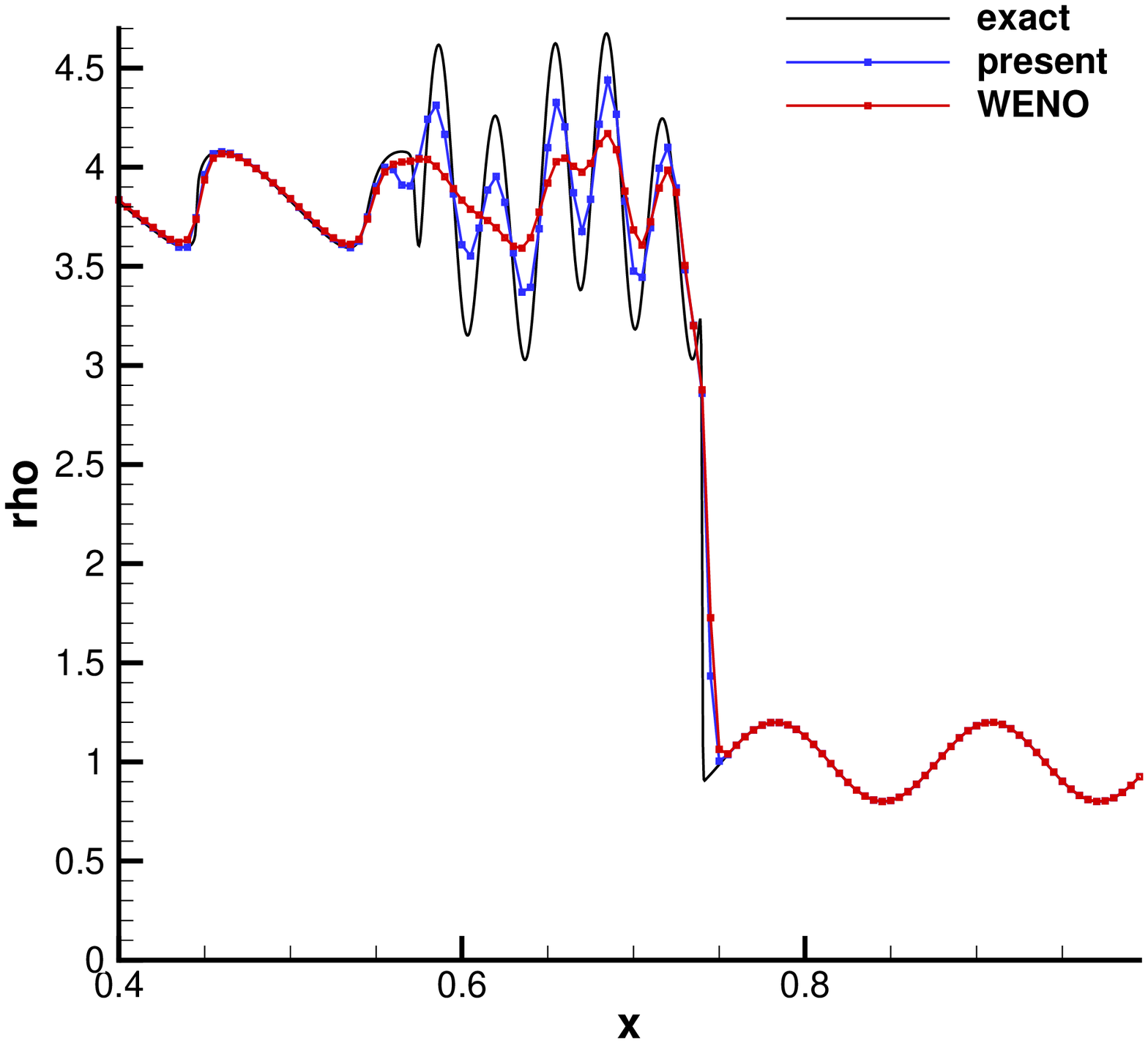}}
	{\centering\includegraphics[scale=0.35,trim={0.5cm 0.5cm 0.5cm 0.5cm},clip]{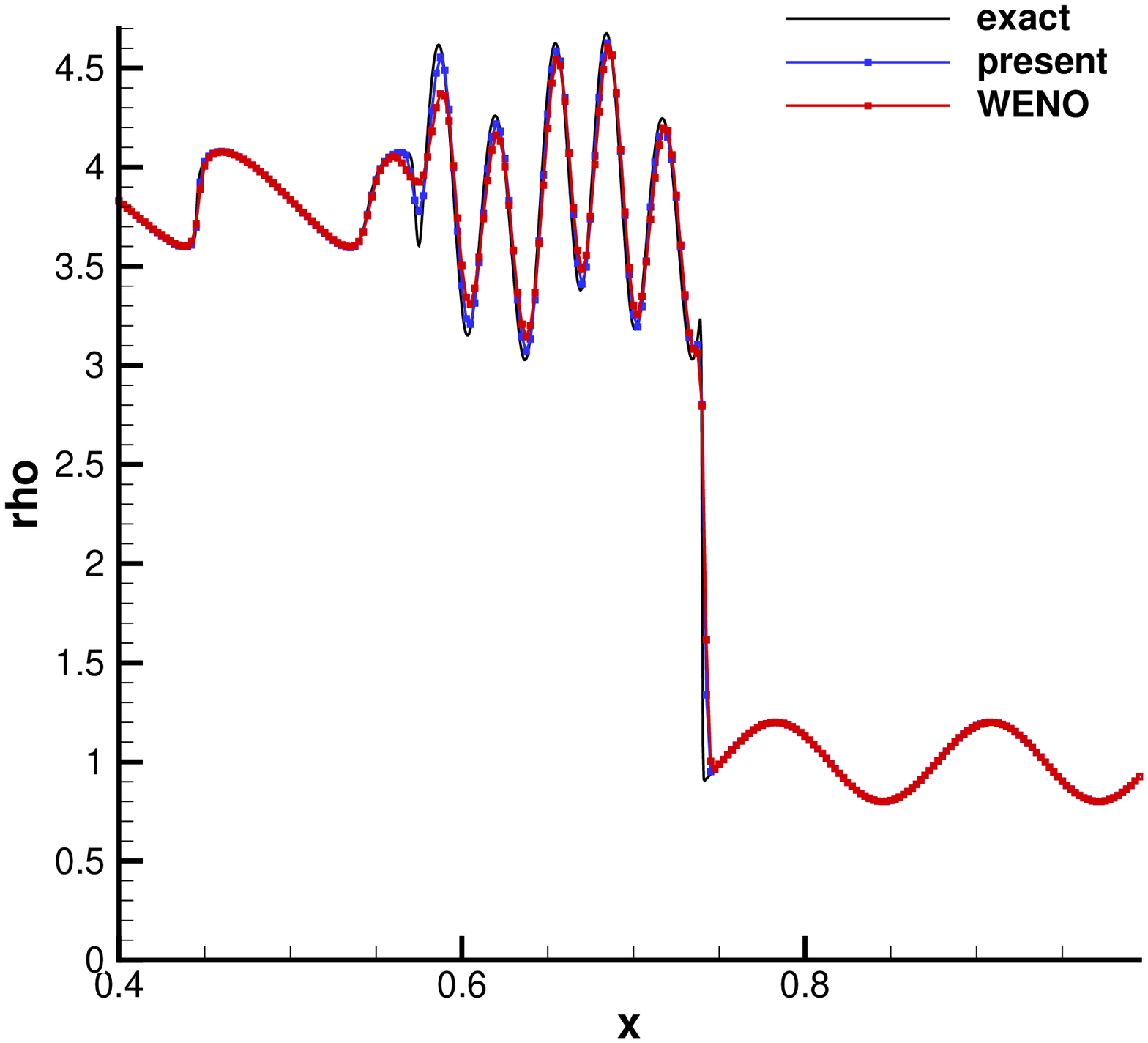}}
	\protect\caption{Numerical results of shock/density wave interaction problem. Comparisons are made between the proposed scheme and WENO scheme against the exact solution. The results with 200 mesh cells are shown in the left panel and those with 400 mesh cells in the right panel.
		\label{fig:shockT}}	
\end{figure}

A high-order shock capturing scheme is expected to be able to solve problems containing shocks and complex smooth flow features. Here, we simulate the case proposed by \cite{shu_eno2}, which serves as a good model to mimic the interactions in compressible turbulence. In this case, a Mach 3 shock wave interacts with a density disturbance and generates a flow field containing both smooth structures and discontinuities. The initial condition is set as
\begin{equation}
(\rho_{0},\ u_{0},\ p_{0})=\left\{
\begin{array}{ll}
\left(3.857148,\ 2.629369,\ 10.333333\right), \ &\mathrm{if}\ 0 \leq x \leq 0.1,\\
\left(1+0.2\sin\left(50x-25\right),\ 0,\ 1\right), \ &\mathrm{otherwise}.
\end{array}
\right.
\end{equation}
The numerical solutions at $t=0.18$ computed on 200 and 400 mesh elements are shown in Fig.~\ref{fig:shockT}, where the reference solution plotted by the solid line is computed by the classical 5th-order WENO scheme with 2000 mesh cells. It is obvious that the proposed scheme resolves better density perturbations and captures  more accurately the peak of the waves, especially for the case of coarse mesh.

To further demonstrate the high resolution property of the new scheme, another shock density wave interaction problem is tested here. The initial condition is specified similarly as \cite{TitarevToro} 
\begin{equation}
(\rho_{0},\ u_{0},\ p_{0})=\left\{
\begin{array}{ll}
\left(1.515695,\ 0.523346,\ 1.805\right), \ &\mathrm{if}\  x \leq -4.5,\\
\left(1+0.1\sin(10 \pi x),\ 0,\ 1\right), \ &\mathrm{otherwise}.
\end{array}\right.
\end{equation}
This test involves waves of higher wavenumber thus is a better model for shock-turbulence flow containing wider scale of waves. 

\begin{figure}
	{\centering\includegraphics[scale=0.35,trim={0.5cm 0.5cm 0.5cm 0.5cm},clip]{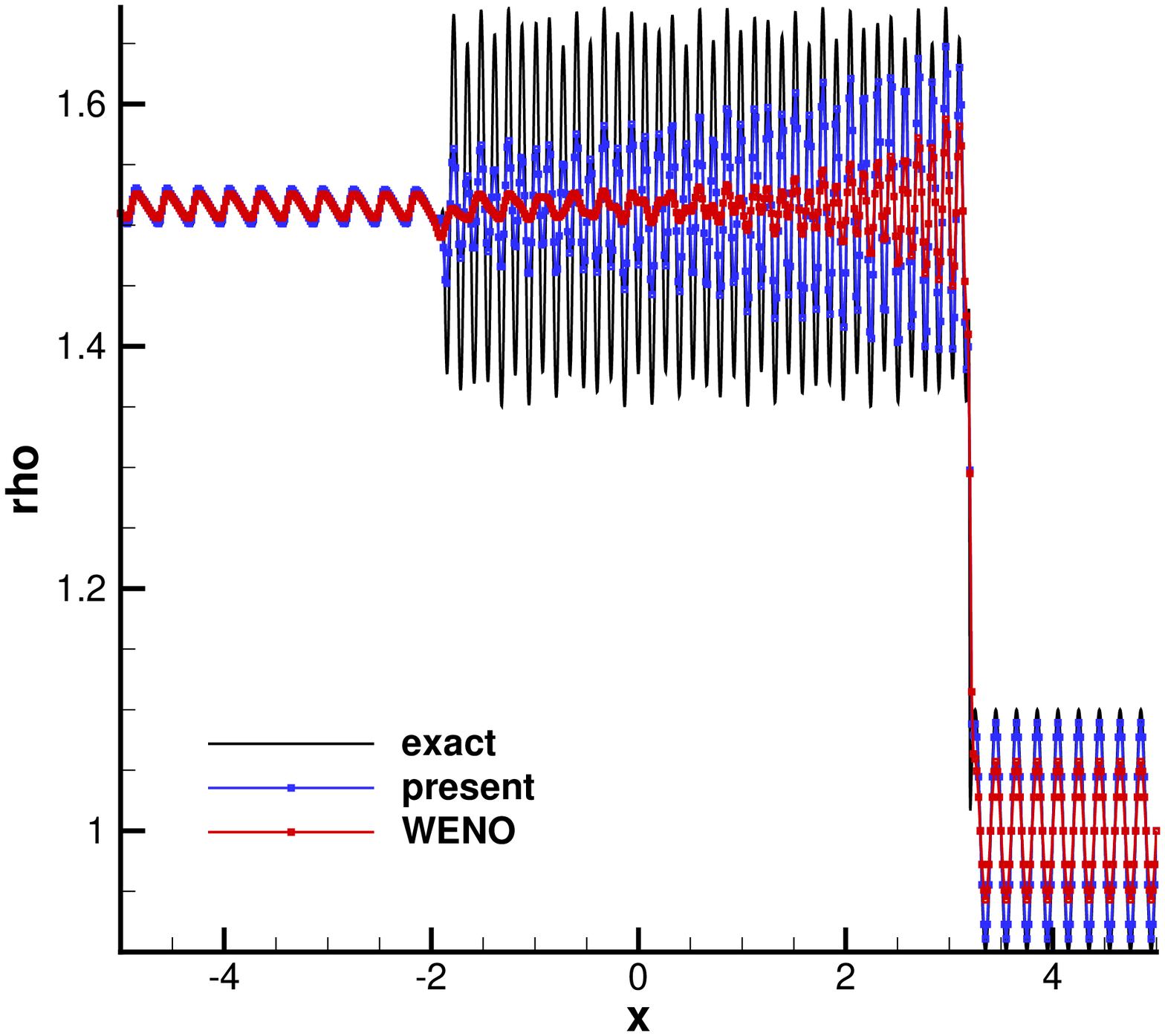}}
	{\centering\includegraphics[scale=0.35,trim={0.5cm 0.5cm 0.5cm 0.5cm},clip]{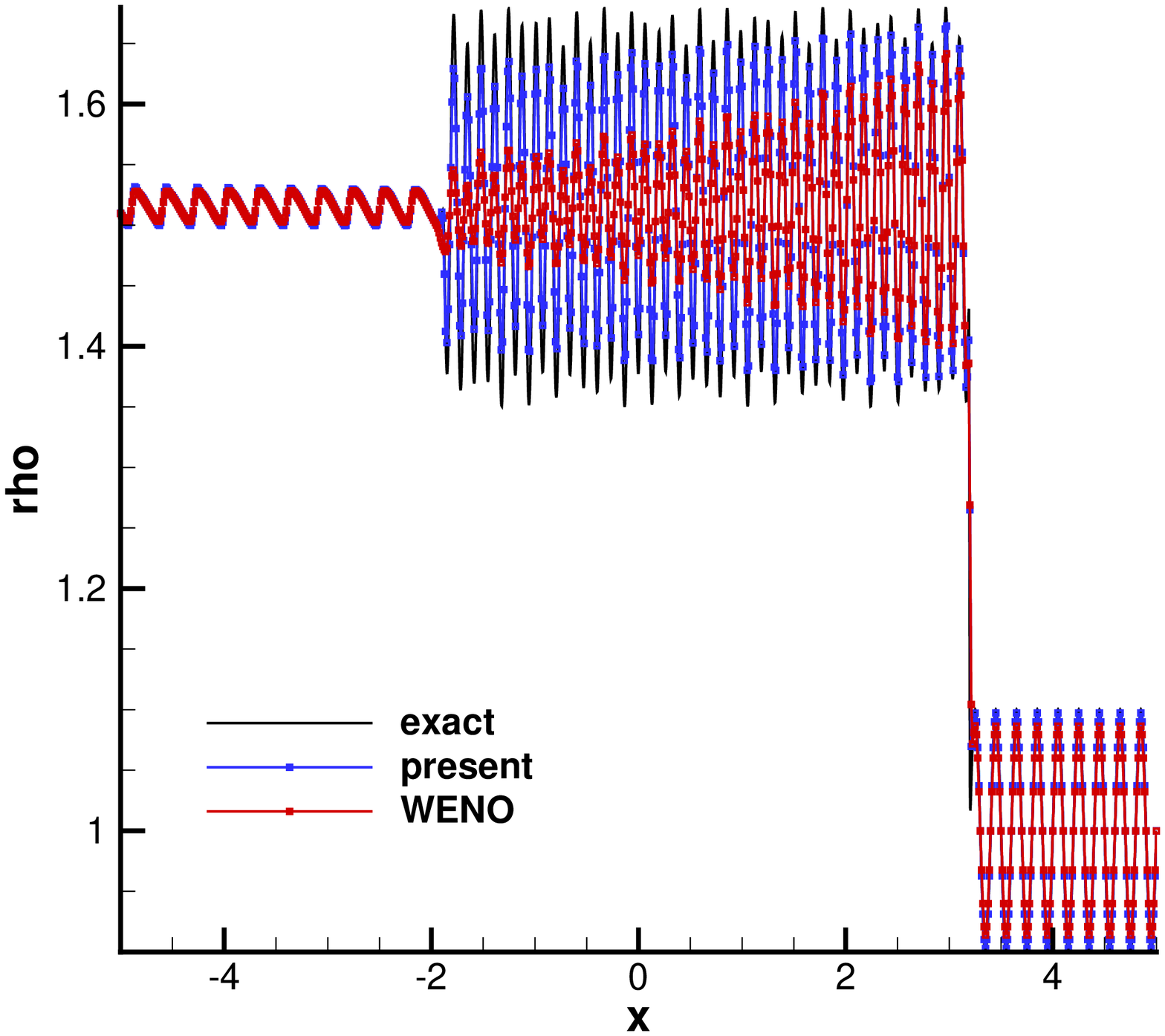}}
	\protect\caption{Numerical results of another shock/density wave interaction problem. The results with 600 mesh elements are shown in the left panel and those with 800 mesh elements in the right panel.
		\label{fig:TT}}	
\end{figure}

\begin{figure}
	{\centering\includegraphics[scale=0.35,trim={0.5cm 0.5cm 0.5cm 0.5cm},clip]{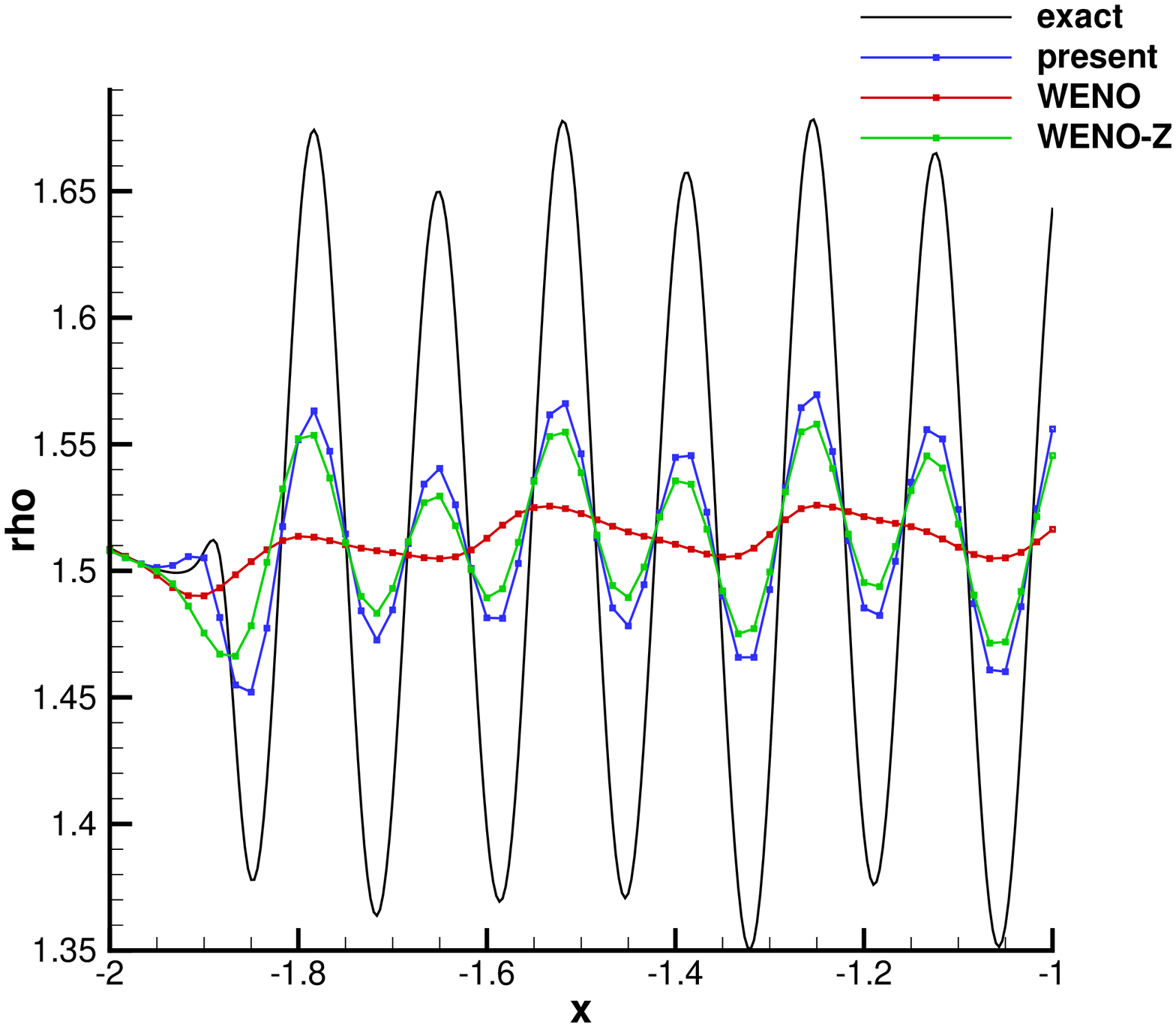}}
	{\centering\includegraphics[scale=0.35,trim={0.5cm 0.5cm 0.5cm 0.5cm},clip]{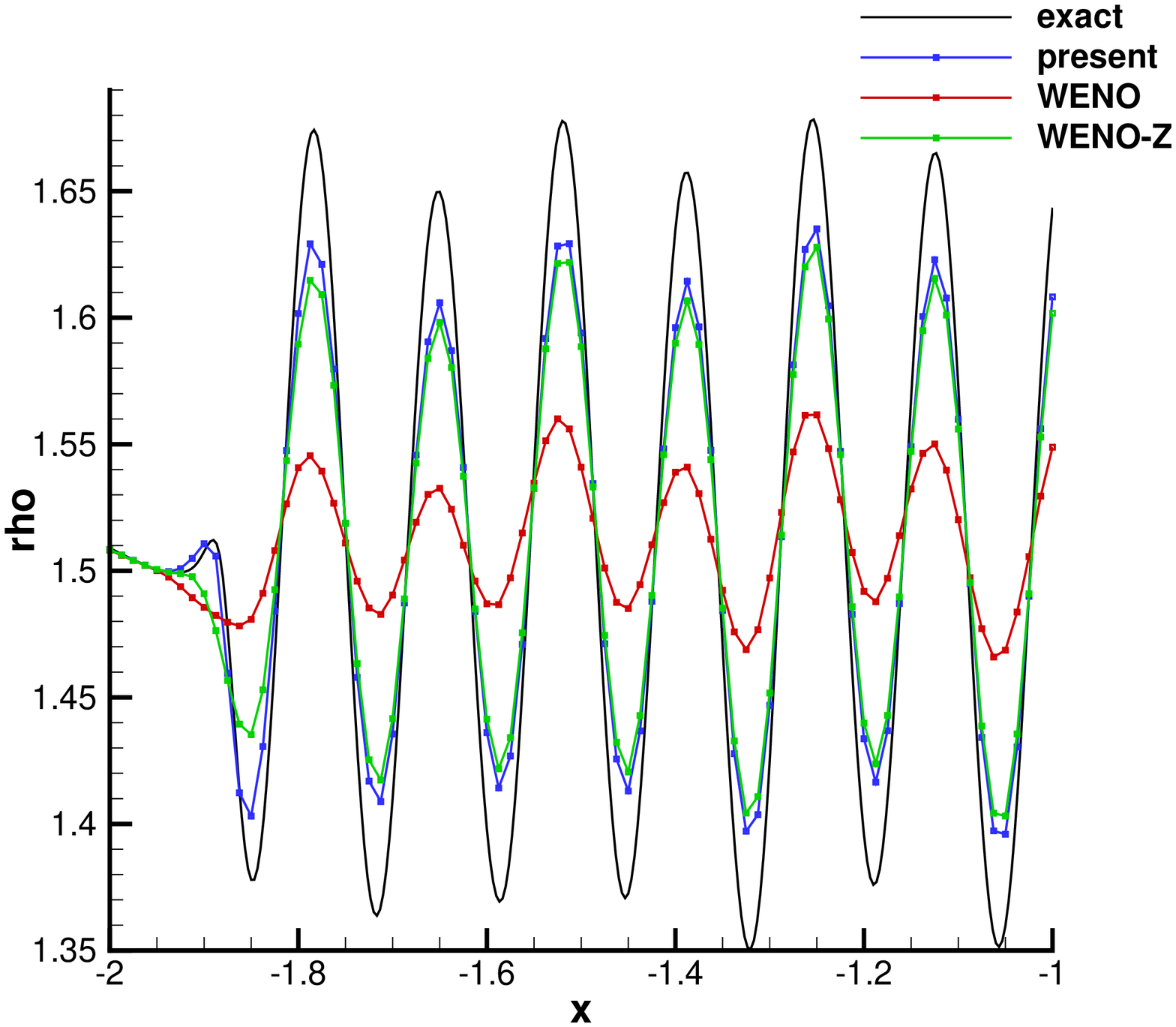}}
	\protect\caption{The same as the Fig.~\ref{fig:TT}, but for a zoomed region of high-wavenumber perturbations.
		\label{fig:TT_L}}	
\end{figure}

The computation is carried out up to $t=5.0$. The numerical solutions with 600 and 800 cells are shown in Fig.~\ref{fig:TT}, where a comparison with the original WENO scheme is included. It can be seen that with a relatively coarse mesh, the flow structures from WENO are largely smeared out, while the new scheme has ability to capture small-scale waves with high accuracy. We further illustrate the performance of the new scheme in comparison with WENO-Z scheme in Fig.~\ref{fig:TT_L} which shows a zoomed region for density perturbation. It is obvious that the new scheme shows the best solution in maintaining the amplitudes of density waves. It reveals that the new scheme has high resolution property to solve complex flow features and is promising to be employed for shock-turbulence interaction problems.

\subsection{Two interacting blast waves}
This benchmark test involves multiple interactions of strong shocks and rarefaction waves, and has been used as a challenging problem to shock capturing schemes since it was introduced in \cite{blast}. The initial distribution is given by
\begin{equation}
(\rho_{0},\ u_{0},\ p_{0})=\left\{
\begin{array}{lll}
\left(1,\ 0,\ 1000\right), \ &\mathrm{if}\ 0 \leq x <0.1,\\
\left(1,\ 0,\ 0.01\right), \ &\mathrm{if}\ 0.1 \leq x <0.9, \\
\left(1,\ 0,\ 100\right),  \ &\mathrm{if}\ 0.9 \leq x <1.
\end{array}\right.
\end{equation}

\begin{figure}
	\begin{centering}
	{\centering\includegraphics[scale=0.35,trim={0.5cm 0.5cm 0.5cm 0.5cm},clip]{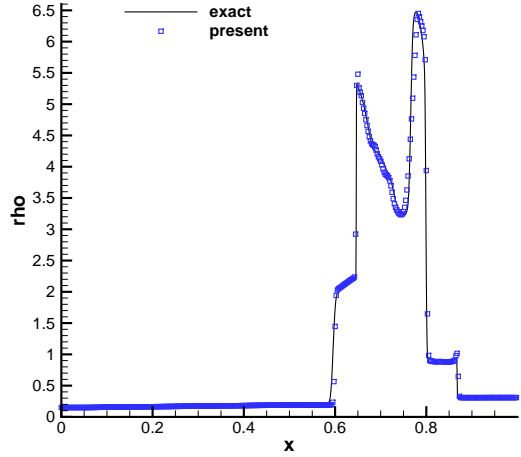}}
	\protect\caption{Numerical results (density field) of two interacting blast waves  at t = 0.038 with 400 cells. \label{fig:blast}}
	\end{centering}	
\end{figure}

Reflective boundary conditions are imposed at the two ends of computational domain. 
Two blast waves are generated by the initial jumps in pressure and then develop complex structures through violent interactions. 
We use $400$ mesh cells as usually found in the literature for this test problem.
The numerical result of density at time $t = 0.038$ is depicted in Fig.~\ref{fig:blast} against a reference solution 
obtained with WENO scheme on  a very fine mesh.
The solution produced by various WENO schemes on $400$-cell mesh can be found in many published works, such as  \cite{jiang96,wenoz} where contact discontinuities are smeared significantly, especially the left-most contact discontinuity around $x\simeq 0.6$. On the contrary the proposed scheme shows overall better resolution and can capture the left-most contact discontinuity  with only three points. 

\section{Conclusion remarks \label{sec:conclusion}}

A novel fifth order shock capturing scheme named P4-THINC-BVD is proposed in the context of finite volume method to solve compressible flows that have both smooth and discontinuous solutions of wide-band scales. In P4-THINC-BVD, a two-stage cascade BVD algorithm is designed to get high-fidelity solutions for both smooth and discontinuous solutions. The present scheme employs the linear fifth order upwind-biased scheme and the THINC functions of adaptive steepness as the reconstruction candidates. 
The final reconstruction function is selected from these candidates through a two-stage cascade BVD algorithm that minimizes the jumps of the reconstructed values at cell boundaries. 

As shown in spectral analysis, if a linear scheme with high order polynomial is used together with other monotonicity reinforced schemes, like TVD or THINC schemes, the BVD algorithm can retrieve the underlying low-dissipation linear scheme for all wave numbers in smooth solutions. The benchmark tests verify that with effectively suppressed numerical oscillations and significantly reduced numerical dissipation, the proposed P4-THINC-BVD is capable of capturing sharp discontinuities and resolving small-scale flow structures with substantially improved solution quality in comparison with other existing methods. The P4-THINC-BVD scheme as well as the basic idea presented in this paper provides an innovative and practical alternative approach for spatial reconstructions in high order finite volume method to solve hyperbolic conservative systems.

\section*{Acknowledgment}
This work was supported in part by the fund from JSPS (Japan Society for the Promotion of Science) 
under Grant Nos. 15H03916, 15J09915 and 17K18838. 

\begin{figure}
	\begin{centering}
		\subfigure[Minmod]{\centering\includegraphics[scale=0.28,trim={0.9cm 0.9cm 0.9cm 0.9cm},clip]{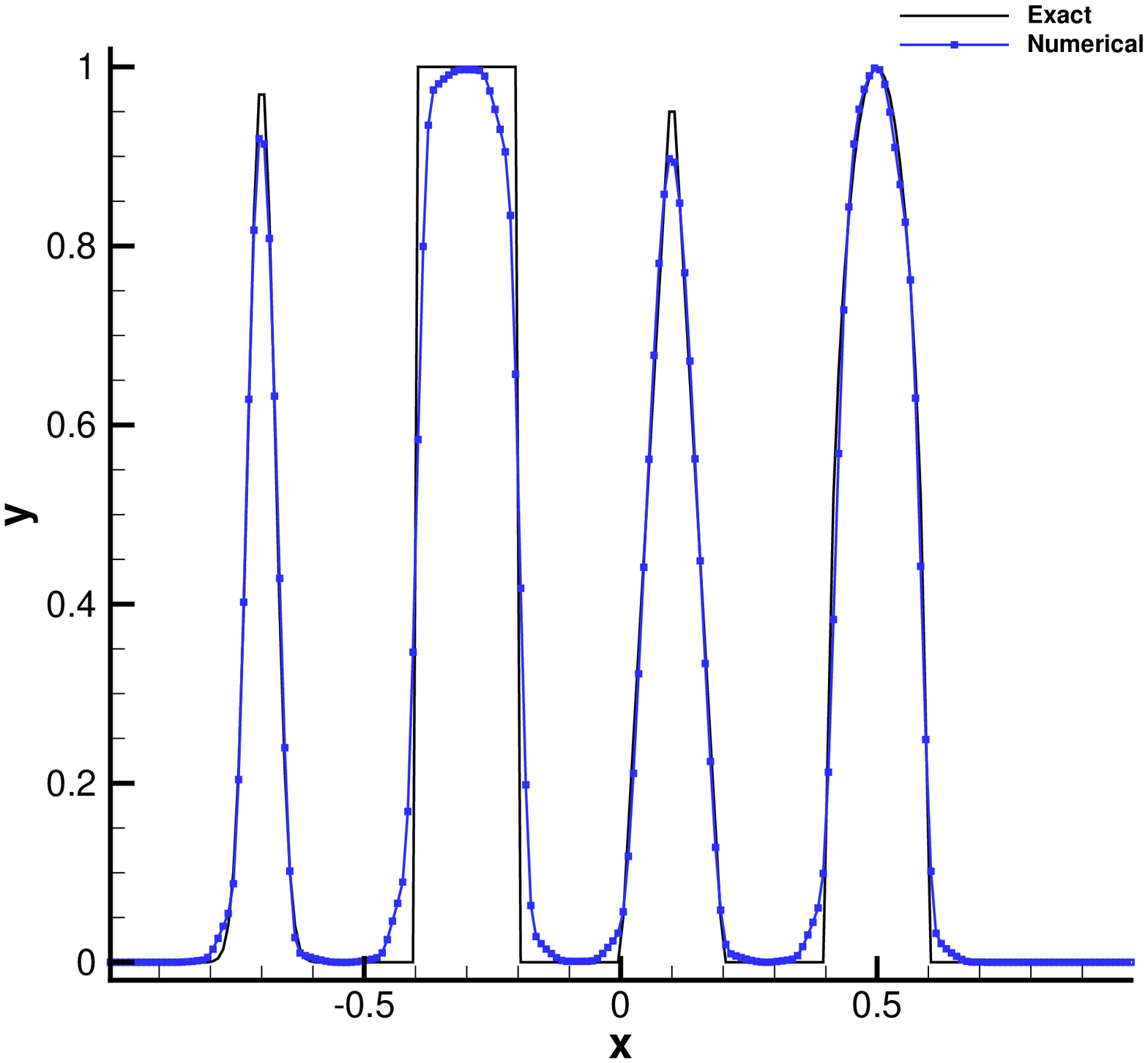}} 
		\subfigure[Van Leer]{\centering\includegraphics[scale=0.28,trim={0.9cm 0.9cm 0.9cm 0.9cm},clip]{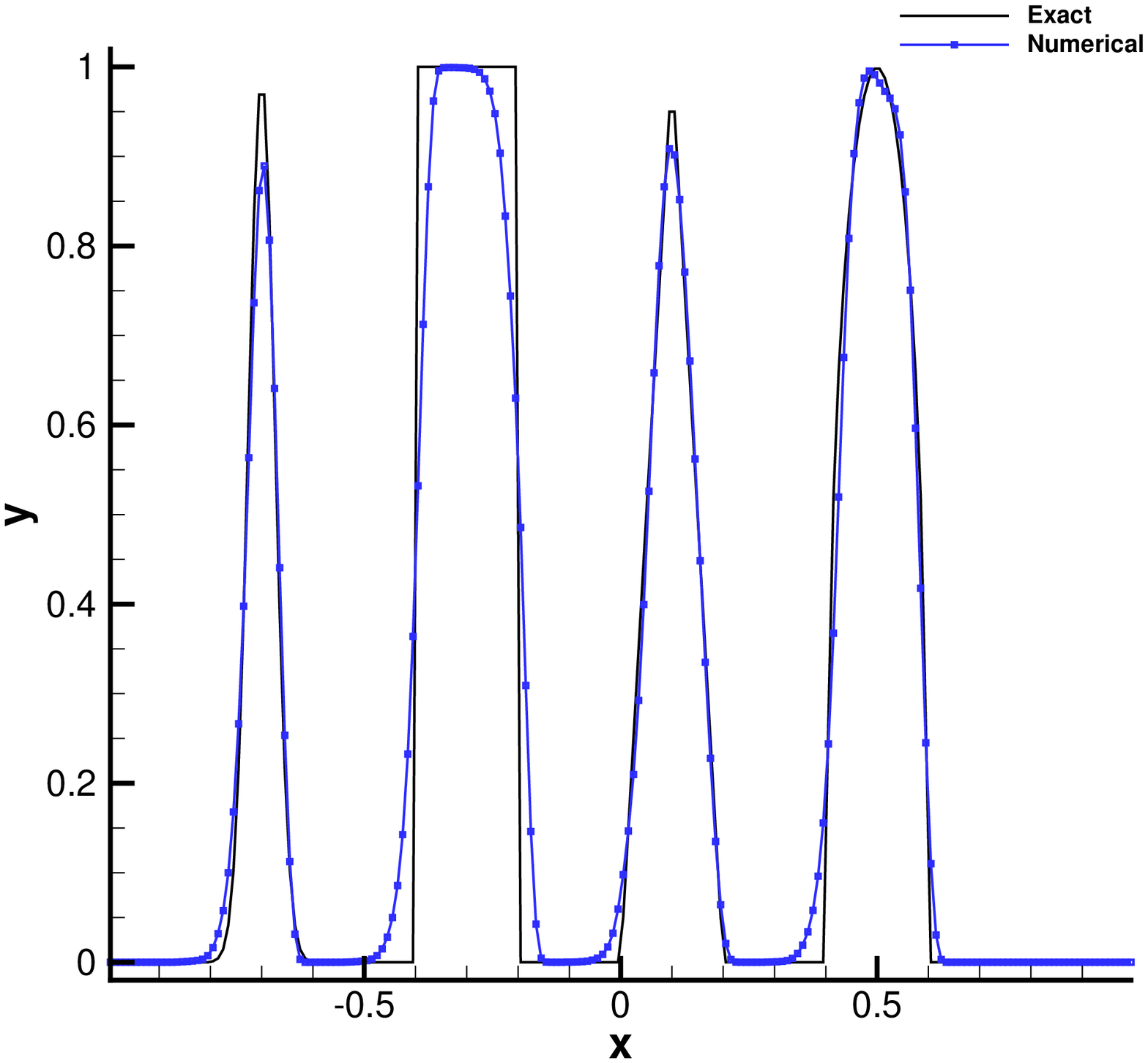}} 
		\subfigure[Superbee]{\centering\includegraphics[scale=0.28,trim={0.9cm 0.9cm 0.9cm 0.9cm},clip]{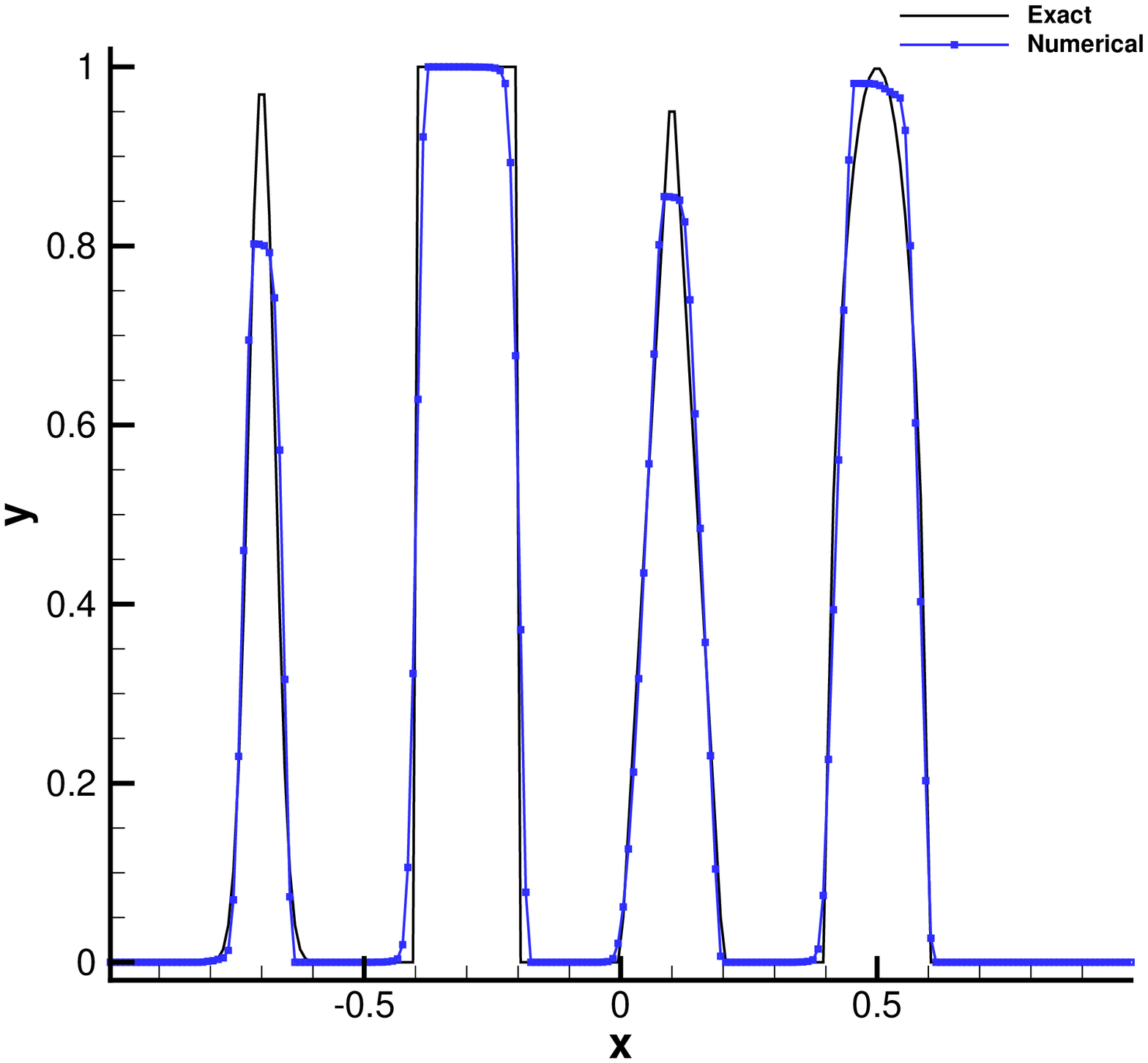}}
		\subfigure[THINC($\beta_{s}$)]{\centering\includegraphics[scale=0.28,trim={0.9cm 0.9cm 0.9cm 0.9cm},clip]{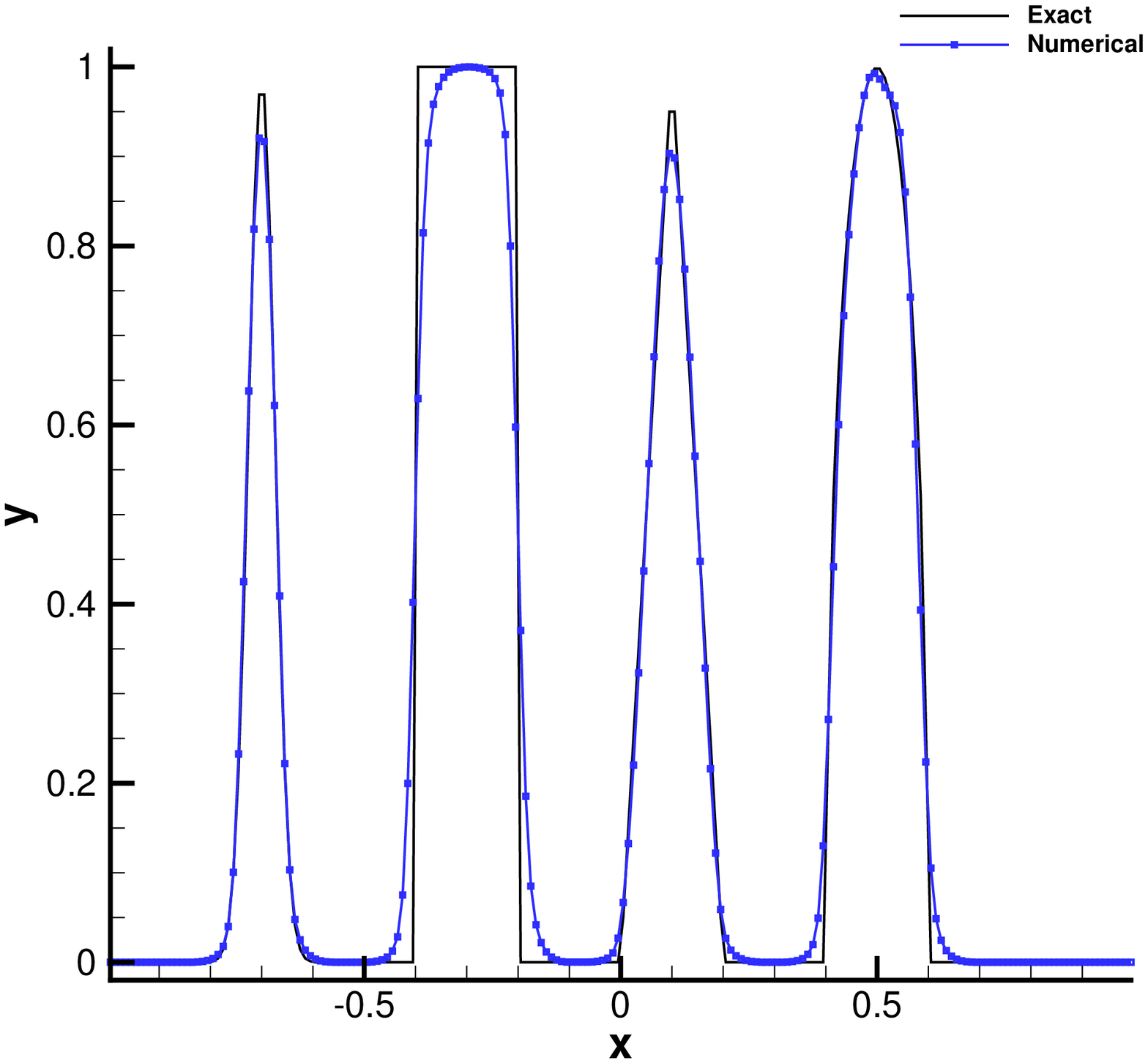}}
		\protect\caption{Numerical results for advection of complex waves  at $t=2.0$ with 200 mesh cells. Displayed are  solutions of the first-stage BVD presented in \S2.2.3 where the  minmod (a), van Leer (b), superbee (c) and THINC($\beta_{s}$) (d)  schemes are used respectively as the candidate interpolants together with the polynomial of degree four.
			\label{fig:appendix}}	
	\end{centering}
\end{figure}

\begin{figure}[h]
	\begin{center}
		\subfigure[Minmod]{\centering\includegraphics[scale=0.35,trim={0.0cm 0.0cm 0.0cm 0.0cm},clip]{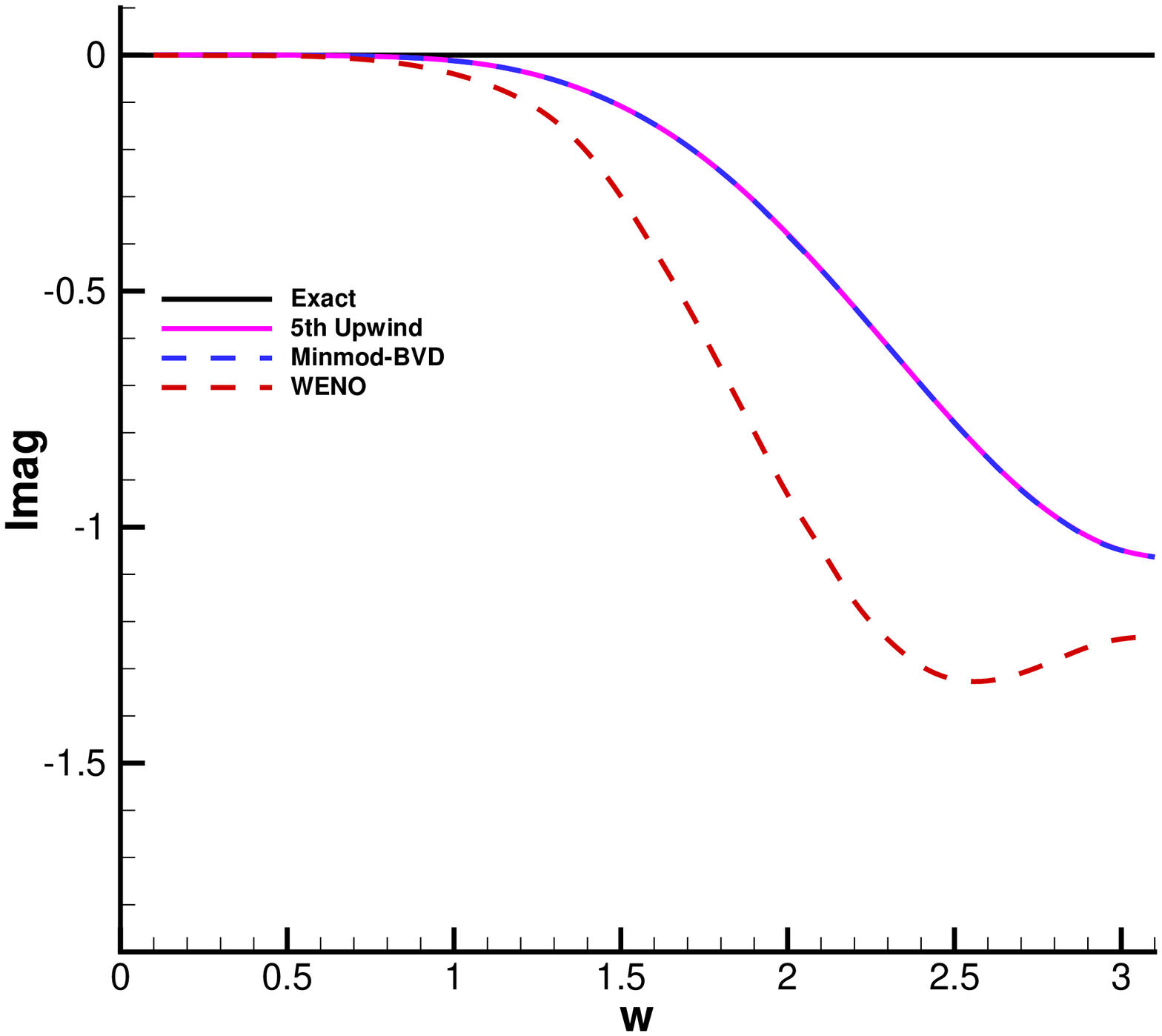}}
		\subfigure[Van Leer]{\centering\includegraphics[scale=0.35,trim={0.0cm 0.0cm 0.0cm 0.0cm},clip]{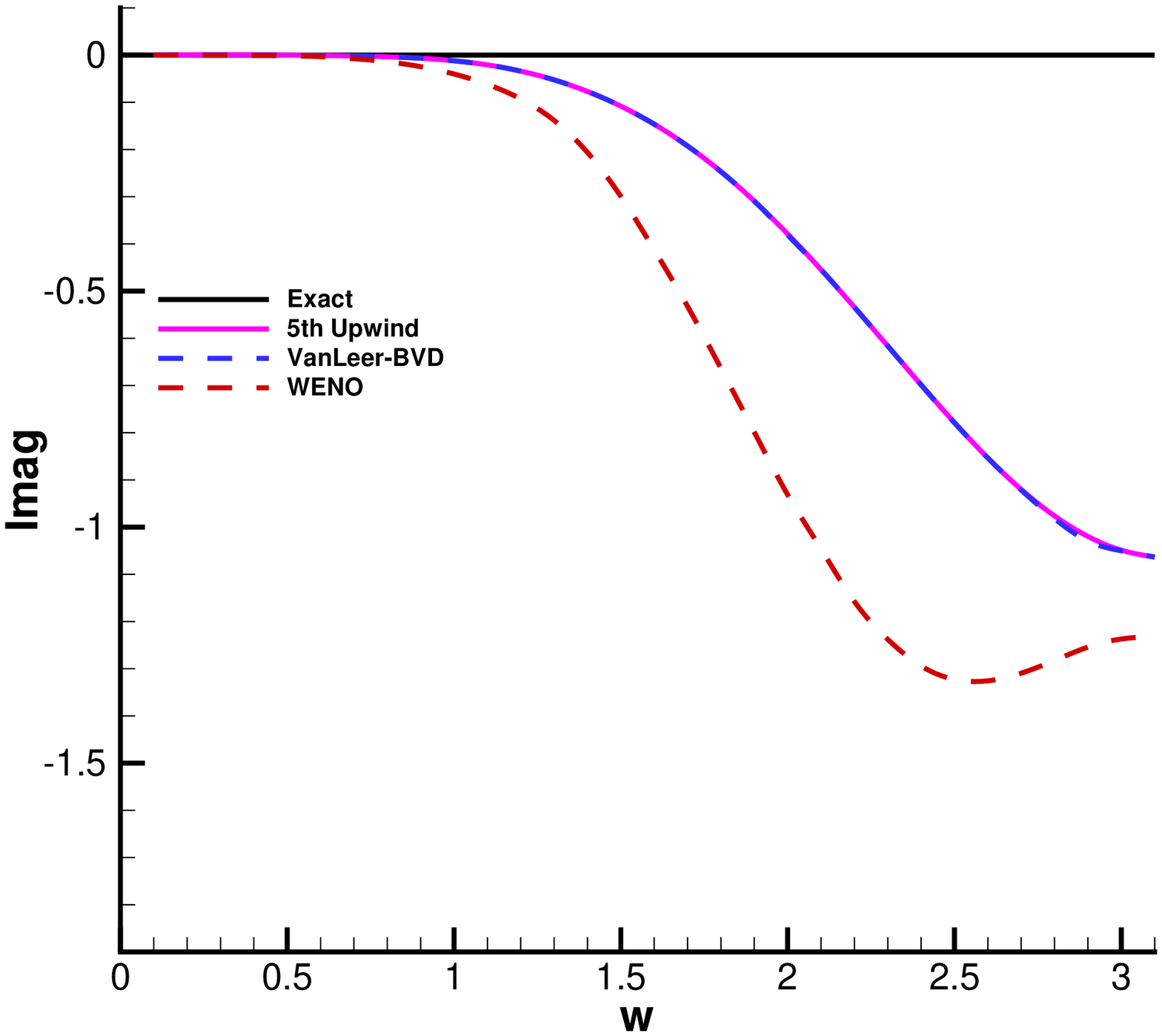}}
		\subfigure[Superbee]{\centering\includegraphics[scale=0.35,trim={0.0cm 0.0cm 0.0cm 0.0cm},clip]{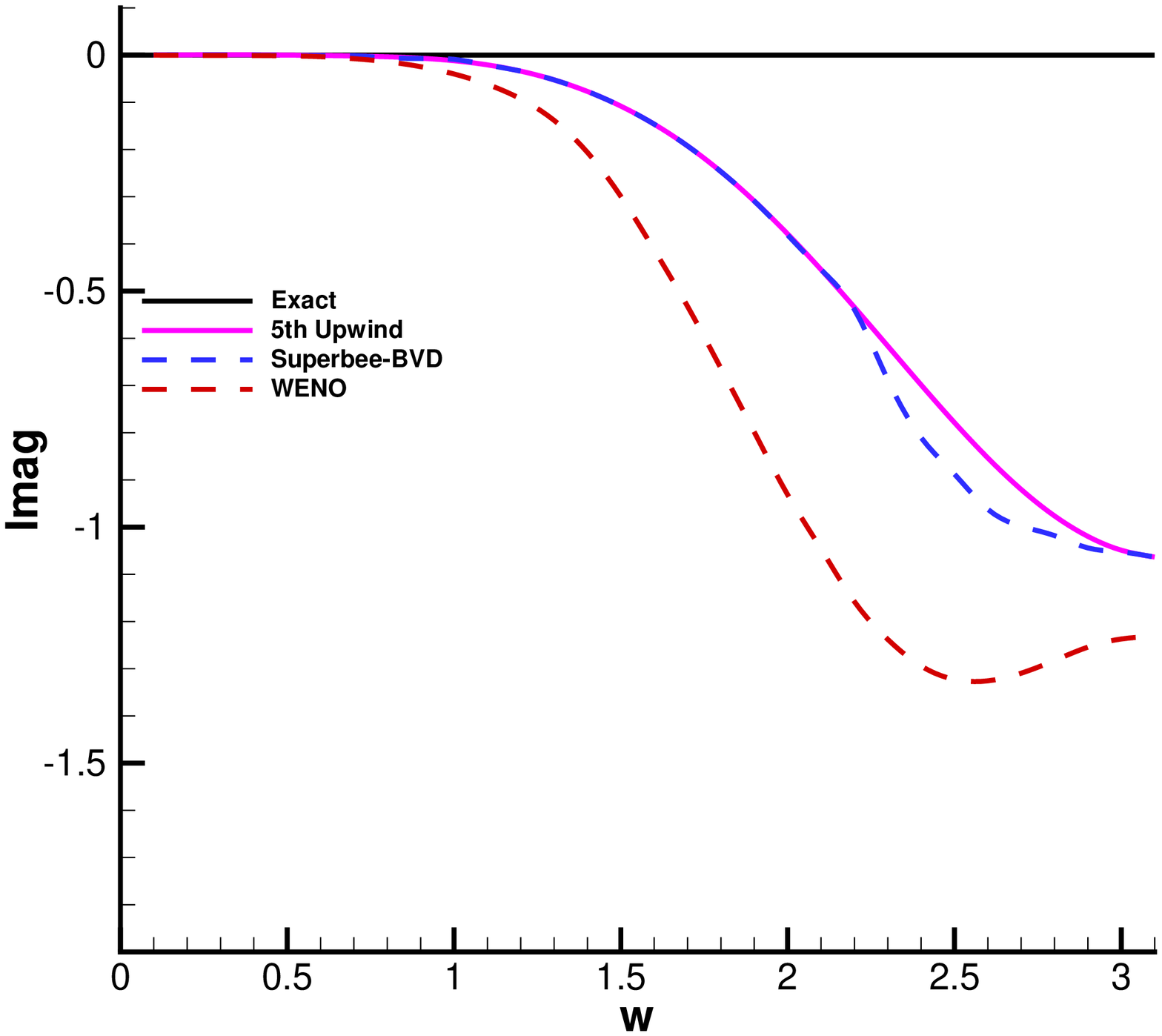}}
		\subfigure[THINC($\beta_{s}$)]{\centering\includegraphics[scale=0.35,trim={0.0cm 0.0cm 0.0cm 0.0cm},clip]{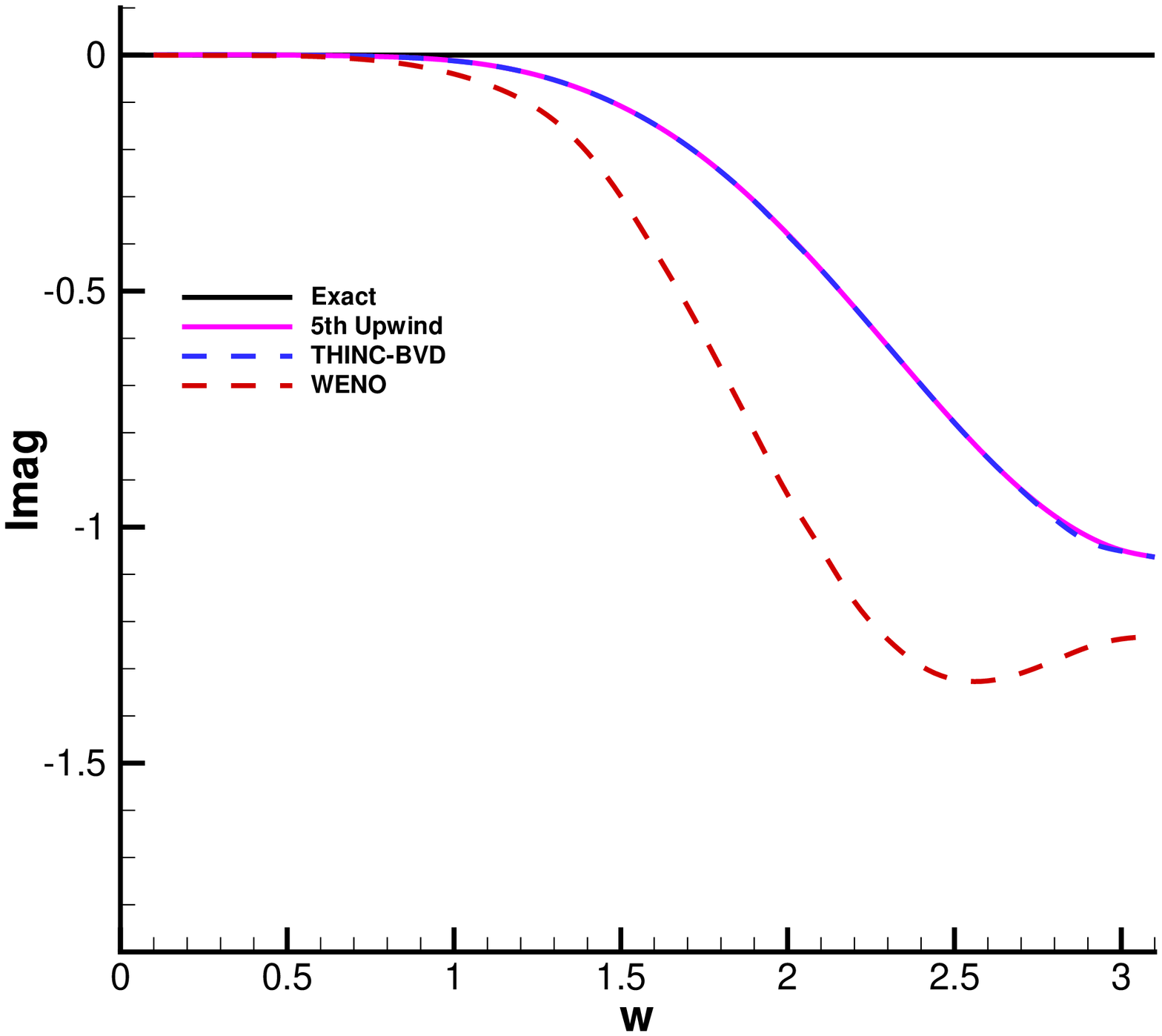}}
		\protect\caption{Comparison of the spectral properties of the BVD schemes and the WENO scheme. Displayed are the imaginary parts of the modified wavenumber calculated by the first-stage BVD presented in \S2.2.3 where the  minmod (a), van Leer (b), superbee (c) and THINC($\beta_{s}$) (d)  schemes are used respectively as the candidate interpolants together with the polynomial of degree four.
			\label{fig:ADRappendix}}
	\end{center}	
\end{figure} 
\section*{Appendix A. Oscillation-free solution of BVD method with 4th-degree polynomial and TVD  candidate interpolants}
In this appendix, we show that the  BVD algorithm in stage (I)  described in \S2.2.3 also effectively suppresses numerical oscillation when the THINC interpolant is replaced by other existing TVD schemes. Numerical results for the advection test of complex waves in subsection \ref{sec:complex} are presented in Fig.~\ref{fig:appendix}. It can be seen that numerical solutions are essentially non-oscillatory. Thus, BVD algorithm provides a new general strategy that are not only limited to the THINC function but also admits other monotonicity-reinforced schemes as the candidate reconstruction functions  to suppress numerical oscillations caused by high order interpolations. 

\section*{Appendix B. Spectral analysis of the BVD method with 4th-degree polynomial and TVD candidate interpolants}
We examine the spectral properties of the BVD method that uses other existing  monotonic interpolations instead of the THINC function.  Fig.~\ref{fig:ADRappendix} shows the numerical dissipations (imaginary parts of the modified wavenumbers) of the  BVD algorithm in stage (I)  presented in \S2.2.3 with different TVD type candidate schemes. It is observed that the BVD limiting projection which combines the high-order linear upwind scheme with other nonlinear oscillation-suppressing schemes can also preserve the spectral properties of the underlying linear upwind scheme with high-order polynomials. It implies a great advantage of the BVD method over the schemes based on WENO methodology.

\clearpage{}



\begin{thebibliography}{99}
\bibitem{harten-tvd}A. Harten, High resolution schemes for hyperbolic conservation laws, J. Comput. Phys. 49 (1983) 357-393.

\bibitem{sweby84} P.K. Sweby, High resolution schemes using flux-limiters for hyperbolic conservation laws, SIAM J. Num. Anal., 21(1984) 995-1011. 

\bibitem{Van_Leer}B. Van Leer, Towards the ultimate conservative difference scheme. V. A second-order sequel to Godunov's method, J. Comput. Phys. 32 (1979) 101-136.

\bibitem{Harten1}A. Harten, S. Osher, Uniformly high-order accurate non-oscillatory schemes, IMRC Technical Summary Rept. 2823, Univ. of Wisconsin, Madison, WI, May 1985.

\bibitem{Harten2}A. Harten, B. Engquist, S. Osher, S. Chakravarthy, Uniformly high order accurate essentially non-oscillatory schemes III, J. Comput. Phys. 71 (1987) 231-323.

\bibitem{shu_eno1}C.W. Shu, S. Osher, Efficient implementation of essentially non-oscillatory shock capturing schemes, J. Comput. Phys. 77 (1988) 439-471.

\bibitem{shu_eno2}C.W. Shu, S. Osher, Efficient implementation of essentially non-oscillatory shock capturing schemes,II, J. Comput. Phys. 83 (1989) 32-78.

\bibitem{liu94}X.D. Liu, S. Osher, T. Chan, Weighted essentially non-oscillatory schemes, J. Comput. Phys. 115 (1994) 200-212.

\bibitem{jiang96}G.S. Jiang, C.-W. Shu, Efficient implementation of weighted ENO schemes, J. Comput. Phys. 126 (1996) 202-228.

\bibitem{WENOM}A.K. Henrick, T.D. Aslam, J.M. Powers, Mapped weighted essentially non-oscillatory schemes: achieving optimal order near critical points, J. Comput. Phys. 207 (2005) 542-567.

\bibitem{TENO14}S. Zhao, N. Lardjane, I. Fedioun, Comparison of improved finite-difference WENO schemes for the implicit large eddy simulation of turbulent non-reacting and reacting high-speed shear flows, Comput. Fluids 95 (2014) 74-87.

\bibitem{TENO15}M.P. Martin, Shock-Capturing in LES of High-Speed Flows, Center for Turbulence Research Annual Research Briefs, 2000.

\bibitem{wenoz}R. Borges, M. Carmona, B. Costa, W.S. Don, An improved weighted essentially non-oscillatory scheme for hyperbolic conservation laws, J. Comput. Phys. 227 (2008) 3191-3211.

\bibitem{wenoh}Y.Ha, C.H. Kim, Y.J. Lee, J. Yoon, An improved weighted essentially non-oscillatory scheme with a new smoothness indicator, J. Comput. Phys. 232 (2013) 68-86.

\bibitem{wenop}P.Fan, Y. Shen, B. Tian, C. Yang, A new smoothness indicator for improving the weighted essentially non-oscillatory scheme, J. Comput. Phys. 269 (2014) 329-354.

\bibitem{wenozn}F. Acker, R. Borges, B. Costa, An improved WENO-Z scheme, J. Comput. Phys. 313 (2016) 726-753.

\bibitem{wenohu}X.Y. Hu, Q. Wang, N.A. Adams, An adaptive central-upwind weighted essentially non-oscillatory scheme, J. Comput. Phys. 229 (2010) 8952-8965.

\bibitem{TENO}L. Fu, X.Y. Hu, N.A. Adams, A family of high-order targeted ENO schemes for compressible-fluid simulations, J. Comput. Phys. 305 (2016) 333-359.

\bibitem{embedded}B.S. van Lith, J.H. ten Thije Boonkkamp, W.L. IJzerman, Embedded WENO: A design strategy to improve existing WENO schemes, J. Comput. Phys. 330 (2017) 529-549.

\bibitem{Sun}Z. Sun, S. Inaba, F. Xiao, Boundary Variation Diminishing (BVD) reconstruction: A new approach to improve Godunov schemes, J. Comput. Phys. 322 (2016) 309-325.	

\bibitem{xie2017}B. Xie, X. Deng, Z. Sun, F. Xiao, A hybrid pressure-density-based Mach uniform algorithm for 2D Euler equations on unstructured grids by using multi-moment finite volume method, J. Comput. Phys. 335 (2017) 637-663.

\bibitem{deng2018a}X. Deng, S. Inaba, B. Xie, K.M. Shyue, F. Xiao, High fidelity discontinuity-resolving reconstruction for compressible multiphase flows with moving interfaces, J. Comput. Phys. 371 (2018) 945-966. 

\bibitem{deng2018b}X. Deng, B. Xie, F. Xiao, H. Teng, New Accurate and Efficient Method for Stiff Detonation Capturing, AIAA J. (2018) 1-15. 

\bibitem{dengCF}X. Deng, B. Xie, R. Loub{\`e}re, Y. Shimizu, F. Xiao, Limiter-free discontinuity-capturing scheme for compressible gas dynamics with reactive fronts, Comput. Fluids, 171 (2018) 1-14. 

\bibitem{very3}D.S. Balsara, C.W. Shu, Monotonicity prserving WENO schemes with increasingly high-order of accuracy, J. Comput. Phys. 160 (2000) 405-452.

\bibitem{xiao_thinc}F. Xiao, S. Ii, C. Chen, Revisit to the THINC scheme: a simple algebraic VOF algorithm, J. Comput. Phys. 230 (2011) 7086-7092.

\bibitem{xiao_thinc2}F. Xiao, Y. Honma, T. Kono, A simple algebraic interface capturing scheme using hyperbolic tangent function, Int. J. Numer. Methods Fluids 48 (2005) 1023-1040.

\bibitem{books}C. Hirsch, Numerical computation of internal and external flows: The fundamentals of computational fluid dynamics, Butterworth-Heinemann (2007).

\bibitem{adr}S. Pirozzoli, On the spectral properties of shock-capturing schemes, J. Comput. Phys. 219 (2006) 489-497.

\bibitem{sod}G.A. Sod, A survey of several finite difference methods for systems of nonlinear hyperbolic conservation laws, J. Comput. Phys. 27 (1978) 1-31.

\bibitem{TitarevToro}V.A. Titarev, E.F. Toro, Finite-volume WENO schemes for three-dimensional conservation laws, J. Comput. Phys. 201 (2004) 238-260.

\bibitem{blast}P. Woodward, P. Colella, The numerical simulation of two-dimensional fluid flow with strong shocks, J. Comput. Phys. 54 (1984) 115-173.

\bibitem{2dH1}P. Buchm\"{u}ller, C. Helzel, Improved accuracy of high-order WENO finite volume methods on Cartesian grids, J. Sci. Comput. 61 (2014) 343-368.

\bibitem{2dH2}R. Zhang, M. Zhang, C.W. Shu, On the order of accuracy and numerical performance of two classes of finite volume WENO schemes, Commun. Comput. Phys. 9 (2011) 807-827.

\bibitem{veryhigh}G.A. Gerolymos, D. S\'en\'echal, I.Vallet, Very-high-order WENO schemes, J. Comput. Phys. 228 (2009) 8481-8524.

\bibitem{yan}Z.H. Jiang, C. Yan, J. Yu, Efficient methods with higher order interpolation and MOOD strategy for compressible turbulence simulations, J. Comput. Phys. 371 (2018) 528-550.

\bibitem{toro}E.F. Toro, 2013. Riemann solvers and numerical methods for fluid dynamics: a practical introduction. Springer Science \& Business Media.

\bibitem{rie1}C.W. Schulz-Rinne, Classification of the Riemann problem for two-dimensional gas dynamics, SIAM J. Math. Anal 24 (1993) 76-88.

\bibitem{rie2}A. Kurganov, E. Tadmor, Solution of two-dimensional Riemann problems for gas dynamics without Riemann problem solvers,  Numer. Methods Partial Differential Equations 18 (2002) 584-608.

\bibitem{rie_Dumbser}M. Dumbser, O. Zanotti, R. Loub{\`e}re, S. Diot, A posteriori subcell limiting of the discontinuous Galerkin finite element method for hyperbolic conservation laws, J. Comput. Phys. 278 (2014) 47-75.

\bibitem{rie_N1}Y. Ha,  C.H. Kim, Y.J. Lee, J. Yoon, An improved weighted essentially non-oscillatory scheme with a new smoothness indicator, J. Comput. Phys. 232 (2013) 68-86.

\bibitem{rie_N2}P. Buchm\"uller, C. Helzel, Improved accuracy of high-order WENO finite volume methods on Cartesian grids, J. Sci. Comput. 61 (2014) 343-368.

\bibitem{rie_cpc}R. Abedian,  H. Adibi, M. Dehghan, A high-order symmetrical weighted hybrid ENO-flux limiter scheme for hyperbolic conservation laws, Comput. Phys. Comm. 185 (2014) 106-127.

\bibitem{rie_fine}C.Y. Jung, T.B. Nguyen, Fine structures for the solutions of the two-dimensional Riemann problems by high-order WENO schemes, Adv. Comput. Math. (2017) 1-28.

\bibitem{double}P. Woodward, P. Colella, The numerical simulation of two-dimensional fluid flow with strong shocks, J. Comput. Phys. 54 (1984) 115-173.

\bibitem{SVstrong}A. Rault, G. Chiavassa, R. Donat, Shock-vortex interactions at high Mach numbers, J. Sci. Comput. 19 (2003) 347-371

\bibitem{Dumbser39}M. Dumbser, M. K\"aser, V.A. Titarev, E.F. Toro, Quadrature-free non-oscillatory finite volume schemes on unstructured meshes for nonlinear hyperbolic systems, J. Comput. Phys. 226 (2007) 204-243.



\end{thebibliography}
\end{document}